\documentclass{aa}
\usepackage{natbib}
\usepackage{hyperref}
\usepackage[none]{hyphenat}
\usepackage{appendix, caption}

\usepackage{graphicx}
\usepackage{txfonts}
\usepackage[mathscr]{euscript}
\usepackage{color}

\usepackage{longtable}
\usepackage{pdflscape}
\usepackage{color}
\usepackage{pdftexcmds}

\def\galex{{\it GALEX\/}} 
\def\spitzer{{\it Spitzer\/}} 
\def\herschel{{\it Herschel\/}}

\def\otelodeep{\textsf{OTELO-Deep}}

\def\oteloint{\textsf{OTELO$_{\rm Int}$}}
\def\oteloone{{OTELO-I}}

\def\acs606{\textit{F606W}}
\def\acs814{\textit{F814W}}
\def\bandu{\textsf{\textit{u}}}
\def\bandg{\textsf{\textit{g}}}
\def\bandr{\textsf{\textit{r}}}
\def\bandi{\textsf{\textit{i}}}
\def\bandz{\textsf{\textit{z}}}
\def\bandj{\textsf{\textit{J}}}
\def\bandh{\textsf{\textit{H}}}
\def\bandk{\textsf{\textit{K$_{\rm s}$}}}
\def\irac36{${\rm 3.6\mu m}$}
\def\irac45{${\rm 4.5\mu m}$}
\def\irac58{${\rm 5.8\mu m}$}
\def\irac80{${\rm 8.0\mu m}$}
\def\mips24{${\rm 24\mu m}$}
\def\pep100{${\rm 100\mu m}$}
\def\pep160{${\rm 160\mu m}$}
\def\her250{${\rm 250\mu m}$}
\def\her350{${\rm 350\mu m}$}
\def\her500{${\rm 500\mu m}$}
\def\irac{{\rm IRAC\/}}
\def\mips{{\rm MIPS\/}}

\def\deep{\hbox{DEEP2}}
\def\otelo{\hbox{OTELO}}

\def\acs{\hbox{HST-ACS}}
\def\cfhtls{\hbox{CFHTLS}}
\def\wirds{\hbox{WIRDS}}
\def\pep{\hbox{PEP}}

\def\zred{{z}}
\def\zs{z$_{\rm\, DEEP2}$}
\def\zp{z$_{\rm\, phot}$}

\def\zotelo{z$_{\rm\, OTELO}$}

\def\g2{\texttt{GALAPAGOS-2}}

\def\ha{H$\alpha$}
\def\hb{H$\beta$}

\def\ni{[N\,I]\,$\lambda$5199}
\def\nii{[N\,II]\,$\lambda$6548,6583}

\def\oii{[O\,II]\,$\lambda$3726,3729}
\def\oiii{[O\,III]\,$\lambda$4959,5007}
\def\oiiis{[O\,III]}
\def\oiiir{[O\,III]\,$\lambda$5007}
\def\oiiib{[O\,III]\,$\lambda$4959}
\def\oiiia{[O\,III]\,$\lambda$4363}

\def\sii{[S\,II]\,$\lambda$6717,6731}

\def\heiihb{He\,II\,$\lambda$4686}

\begin{document} 

\title{The OTELO survey}
\subtitle{A case study of \oiii\ emitters at $\langle$z$\rangle$ = 0.83}
\author{\'Angel Bongiovanni\inst{1,2,3,4}
\and Marina Ram\'on-P\'erez\inst{2,3}
\and Ana Mar\'ia P\'erez Garc\'ia\inst{5,4}
\and Miguel Cervi\~no\inst{5,2,7}
\and Jordi Cepa\inst{2,3,4}
\and Jakub Nadolny\inst{2,3}
\and Ricardo P\'erez Mart\'inez\inst{6,4}
\and Emilio J. Alfaro\inst{7}
\and H\'ector O. Casta\~neda\inst{13}
\and Bernab\'e Cedr\'es\inst{2,3}
\and Jos\'e A. de Diego\inst{12}
\and Alessandro Ederoclite\inst{11,4}
\and Mirian Fern\'andez-Lorenzo\inst{7}
\and Jes\'us Gallego\inst{9}
\and J. Jes\'us Gonz\'alez\inst{12}
\and J. Ignacio Gonz\'alez-Serrano\inst{8,4}
\and Maritza A. Lara-L\'opez\inst{18}
\and Iv\'an Oteo G\'omez\inst{16,17}
\and Carmen P. Padilla Torres\inst{21}
\and Irene Pintos-Castro\inst{15}
\and Mirjana Povi\'c\,\inst{14,7}
\and Miguel S\'anchez-Portal\inst{1,4}
\and D. Heath Jones\inst{19}
\and Joss Bland-Hawthorn\inst{20}
\and Antonio Cabrera-Lavers\inst{10}
}

\institute{Instituto de Radioastronom\'ia Milim\'etrica (IRAM), Av. Divina Pastora 7, N\'ucleo Central, E-18012 Granada, Spain 
\and Instituto de Astrof\'isica de Canarias (IAC), E-38200 La Laguna, Tenerife, Spain 
\and Departamento de Astrof\'isica, Universidad de La Laguna (ULL), E-38205 La Laguna, Tenerife, Spain 
\and Asociaci\'on Astrof\'isica para la Promoci\'on de la Investigaci\'on, Instrumentaci\'on y su Desarrollo, ASPID, E-38205 La Laguna, Tenerife, Spain 
\and Centro de Astrobiolog\'ia (CSIC/INTA), E-28692, ESAC Campus, Villanueva de la Ca\~nada, Madrid, Spain 
\and ISDEFE for European Space Astronomy Centre (ESAC)/ESA, P.O. Box 78, E-28690, Villanueva de la Ca\~nada, Madrid, Spain 
\and Instituto de Astrof\'isica de Andaluc\'ia, CSIC, E-18080, Granada, Spain  
\and Instituto de F\'isica de Cantabria (CSIC-Universidad de Cantabria), E-39005 Santander, Spain 
\and Departamento de F\'isica de la Tierra y Astrof\'isica \& Instituto de F\'isica de Part\'iculas y del Cosmos (IPARCOS), Facultad de CC. F\'isicas, Universidad Complutense de Madrid, E-28040, Madrid, Spain 
\and Grantecan S.~A., Centro de Astrof\'isica de La Palma, Cuesta de San Jos\'e, E-38712 Bre\~na Baja, La Palma, Spain 
\and Universidade de S\~ao Paulo, Instituto de Astronomia, Geof\'isica e Ci\^encias Atmosf\'ericas, 05508090 - S\~ao Paulo, SP - Brazil 
\and Instituto de Astronom\'ia, Universidad Nacional Aut\'onoma de M\'exico, 04510 Ciudad de M\'exico, Mexico 
\and Departamento de F\'isica, Escuela Superior de F\'isica y Matem\'aticas, Instituto Polit\'ecnico Nacional, 07738 Ciudad de M\'exico, Mexico 
\and Ethiopian Space Science and Technology Institute (ESSTI), Entoto Observatory and Research Center (EORC), Astronomy and Astrophysics Research Division, PO Box 33679, Addis Ababa, Ethiopia 
\and Department of Astronomy \& Astrophysics, University of Toronto, Canada 
\and Institute for Astronomy, University of Edinburgh, Royal Observatory, Blackford Hill, Edinburgh, EH9 3HJ, UK 
\and European Southern Observatory, Karl-Schwarzschild-Str. 2, 85748, Garching, Germany  
\and DARK, Niels Bohr Institute, University of Copenhagen, Lyngbyvej 2, Copenhagen DK-2100, Denmark 
\and English Language and Foundation Studies Centre, University of Newcastle, Callaghan NSW 2308, Australia 
\and Sydney Institute of Astronomy, School of Physics, University of Sydney, NSW 2006, Australia 
\and INAF, Telescopio Nazionale Galileo, Apartado de Correos 565, E-38700 Santa Cruz de la Palma, Spain 
}

\date{Received 15 June 2018 / Accepted 06 January 2020}

  \abstract
   {The OSIRIS Tunable Filter Emission Line Object (\otelo) survey is a very deep, blind exploration of a
selected region of the Extended Groth Strip and is designed for finding emission-line sources (ELSs).
The survey design, observations, data reduction, astrometry, and photometry, 
as well as the correlation with ancillary data used to obtain a final catalogue,  including photo-z 
estimates and a preliminary selection of ELS, were described in a previous contribution.}
   {Here, we aim to determine the main properties and luminosity function (LF) of the \oiiis\ 
ELS sample of \otelo\ as a scientific demonstration of its capabilities, advantages, and 
complementarity with respect to other surveys.
}
   {The selection and analysis procedures of ELS candidates obtained using tunable filter (TF) pseudo-spectra are 
described. We performed simulations in the parameter space of the survey to obtain emission-line detection 
probabilities. Relevant characteristics of \oiiis\ emitters and the LF (\oiiis),
including the main selection biases and uncertainties, are presented.}
   {From 541 preliminary emission-line source candidates selected around \zred=0.8, a total of 184 sources were 
confirmed as \oiiis\ emitters. Consistent with simulations, the minimum detectable line flux and equivalent width (EW) in this
ELS sample are $\sim$5 $\times$ 10$^{-19}$ erg s$^{-1}$ cm$^{2}$ and $\sim$6 \AA, respectively. We are 
able to constrain the faint-end slope ($\alpha = -1.03\pm0.08$) of the observed LF (\oiiis) at a mean redshift of z=0.83. This LF 
reaches values that are approximately ten times lower than those from other surveys. The vast majority (84\%) of the morphologically classified \oiiis\ 
ELSs are disc-like sources, and 87\% of this sample is comprised of galaxies with stellar masses of 
M$_\star$ $<$ 10$^{10}$ M$_{\odot}$.
}
{}

   \titlerunning{\oiiis\ emitters at $\langle$z$\rangle$ = 0.83 in the OTELO survey}
   \authorrunning{Bongiovanni et al.}
   \maketitle

\section{Introduction}
\label{sec:intro}

The analysis of the strength and profile of emission lines in galaxy spectra provides basic information about the kinematics, density
temperature, and chemical composition of the ionised gas, allowing inferences about dust extinction, star-formation
rate, and gas outflows. This enables spectral classification diagnostics between pure star-forming galaxies (SFGs) and hosts 
of active galactic nuclei (AGNs) independently of the galaxy mass. As such,  the detection and characterisation of sources 
with strong nebular emission lines (emission-line sources or ELSs) can provide a wealth of information about galaxy populations 
and the mechanisms that shape the galaxy-evolution process.
 
Dwarf galaxies are not only more numerous than other types of galaxies with intermediate and high masses, but also constitute 
the building blocks of the more massive galaxies within the framework of the hierarchical model of galaxy evolution 
\citep{white91,somerville99,benson03,somerville12}, which would imply an increased number of them with 
redshift. However, little is known about their nature, growth, evolution, or star formation modes; this is particularly true for low- and 
very-low-mass galaxies, that is, those with $\log\,({\rm M}_\star[{\rm M}_{\odot}])\approx$\,7 to 9 \citep{delucia14}. 
Knowledge of the real contribution of dwarf galaxies to the luminosity function (LF) at any epoch is essential for 
understanding various aspects of galaxy evolution \citep{klypin15}, from reionisation phenomena to the fossil records in the Local 
Group. Deep extragalactic surveys for finding faint ELSs, and model-based synthetic 
galaxy catalogues -with their synergies- are fundamental tools in tackling this problem.

From the first photographic atlases (e.g. those of \citealt{sandage61} and \citealt{arp66})  to
the most recent and ongoing surveys, such as SDSS \citep{strauss02}, VVDS-CFDS \citep{lefevre04}, 
zCOSMOS \citep{lilly07}, GAMA \citep{driver09} \deep\ \citep{newman13}, eBOSS \citep{dawson16}, DES \citep{des16}, 
DEVILS \citep{davies18}, hCOSMOS \citep{damjanov18}, VANDELS \citep{mclure18}, and VIPERS \citep{scodeggio18}, 
the number and variety of extragalactic explorations have resulted in a growing understanding of the
key connections between physical processes and fundamental galaxy observables.
In view of this, realistic model-based mock catalogues of galaxies are needed to isolate the net effects and
uncertainties of determined physical galaxy properties on observed galaxies, depending on the techniques used
to gather real galaxy data. Narrow-band (NB) and intermediate-band (IB) imaging stand out among these techniques
for finding ELSs because they do not suffer from the selection biases and can go deeper in flux than classical 
spectroscopy. Surveys such as COMBO-17 \citep{wolf03}, 
ALHAMBRA \citep{moles08}, SHARDS \citep{perezgonzalez13}, J-PAS \citep{benitez14}, CF-HiZELS \citep{sobral15}, 
and the Hyper Suprime-Cam Subaru Strategic Program \citep[HSC-SSP;][and references therein]{hayashi18} have used
this approach to isolate significantly large samples of ELSs in surveys of differently sized comoving volumes.

Even though the NB deep-imaging technique with wide fields of view has been shown to be effective for this 
purpose, in contrast to the conventional spectroscopic surveys because of their proper selection biases and 
required observing time, its sensitivity to low values of equivalent width (EW) is set by the passband of the filter used. 
The line EW of most very faint ELSs is too small to allow isolation of ELS candidates in classical NB surveys or their 
continua is too faint to allow them to be selected as science targets in current spectroscopic surveys.
A compromise solution is the NB scan technique. The OSIRIS Tunable Filter Emission Line Object (\otelo) survey
combines the advantages of blind, deep NB surveys with the typical EW sensitivity of the spectroscopic 
approach for finding dwarf galaxies at low and intermediate redshifts, with a similar performance as that provided
by integrated light surveys for detecting these galaxies in the Local Volume \citep{danieli18}.
In any case, it is foreseeable that future projects such as LSST \citep{lsst09} on very deep NB explorations 
\citep{yoachim19} and 
spectroscopic surveys such as Euclid \citep{laureijs11}, WFIRST \citep{dressler12}, and 
WAVES \citep{driver16} will allow very faint ELS candidates to be  detected in great numbers and completeness.

The aim of this study is to test the potential of \otelo\ for finding dwarf SFGs through the
analysis of an \oiiis\ ELS sample at intermediate redshift in order to constrain its number density
by probing the faint end of the LF(\oiiis),
and prepare the way for studying their intrinsic properties.

The \oiiis\ line originates from the  gas that is  highly ionised by photons from young massive stars in 
the HII regions of galaxies, but is also linked to the narrow-line region (NLR) of AGNs similarly to other 
strong forbidden lines seen in the spectra of low-density plasmas \citep{stasinska07}. Although biased towards
low gas metallicity in high-ionisation conditions, which also generally imply low stellar masses
\citep[][and references therein]{suzuki16}, this emission
line can therefore be used as a  tracer of star formation. Furthermore, the prominence of \oiiis\ emission
among the strongest lines in the optical spectra of these types of sources, as well as its easily
recognisable doublet, makes this emission line the first choice to examine the performance of any 
tunable filter survey dedicated to the search for ELSs.

On the other hand, the observed number density of low-mass ELSs, which 
have shallow gravitational potential wells, is affected by the dominant 
physical processes that regulate star formation triggering or quenching in such galaxies \citep{brough11}. 
The faint end slope of the LF in SFGs -as a way to express that number statistic- and its 
evolution remain as unresolved issues because most recent studies suffer from significant incompleteness at 
luminosities of $\log\,(L\,[L_{\odot}])\lesssim$ 8\,--\,9 and star formation rates (SFRs) lower than 
0.1 ${\rm M}_{\odot}\, yr^{-1}$ \citep{bothwell11}. Evidently, these problems can be progressively solved by 
unbiased, very deep small-area surveys \citep{parsa16}.

This paper is structured as follows. In Section 2, we describe the \otelo\ survey and its main products.
Section 3 outlines the general ELS selection and analysis procedures from \otelo\ data. In Section 4 
we estimate the main characteristic parameters of the survey as a function of the probability of emission-line 
detection based on educated simulations of \otelo\ products. Section 5 is devoted to
obtaining and characterising the final \oiiis\ ELS sample at a mean redshift of $\langle {\rm z}\rangle$ = 0.83 
according to the general procedure. In Section 6, an observed LF is given after summarising the main survey 
biases and uncertainties that affect it. The analysis of this LF and a comparison
with similar data reported in the recent literature are given in Section 7. The last section contains
a compendium of the work presented.
 
For consistency with recent contributions related to the main subject of this paper, we assume a 
standard $\Lambda$-CDM cosmology with $\Omega_\Lambda$=0.7, 
$\Omega_{\rm m}$= 0.3, and H$_0$=70 km s$^{-1}$ Mpc$^{-1}$. All magnitudes are given in the AB system.

\section{The \otelo\ survey data}
\label{sec:otelodesc}


The OTELO survey is an NB imaging survey that scans a spectral window of about 230 \AA\  centred at $\sim$9175 \AA. 
This window is substantially free from strong airglow emission. The survey uses the red tunable filter
(RTF) of the OSIRIS instrument on the 10.4 m 
Gran Telescopio Canarias \cite[GTC,][]{alvarez98} to produce spectral tomography (resolution R$\sim$700) composed 
of 36  slices, each of 12 \AA\ in full width at half maximum (FWHM) and tuned between 9070 and 9280 \AA\ every 6 \AA. 
The net integration time for each slice was 6.6 ks. The spectral sampling adopted is a compromise between 
an intended photometric accuracy of $\sim$20\% in the deblending of the \ha\ and \nii\ emission lines
observed at z$\sim$0.4 and a reasonably short observing time frame.  

The \otelo\ field is located in a selected region of $\sim$56 arcmin$^2$ in the Extended Groth Strip (EGS). 
This region is embedded in Deep Field 3 of the 
Canada--France--Hawaii Telescope Legacy Survey\footnote{\tt http://www.cfht.hawaii.edu/Science/CFHLS} (CFHTLS),
which forms part of the collaborative All-Wavelength EGS International Survey \cite[AEGIS,][]{davis07}.
The design of the \otelo\ RTF scan aims to perform a blind search of mainly extragalactic sources in as many 
disjoint cosmological volumes between redshift z=0.4 and $\sim$6.5 as possible in which strong emission 
lines can be observed in the wavelength range explored. 
The characteristics and data products of \otelo, and the
general demographics of the detected targets are given in the survey presentation paper 
\cite[][hereafter referred to as \oteloone]{OTELO1}. Likewise, that paper contains
a full description of the tunable filters (TFs),  particularly the properties of the RTF.

The coaddition of the 36 NB slices of \otelo\ was used both for a blind source detection in a custom deep 
image (\otelodeep) to create an IB magnitude (\oteloint, 230 \AA\ wide) and to measure the 
fluxes of those sources detected in registered and resampled images in the optical 
\bandu, \bandg, \bandr, \bandi , and \bandz\ bands from \cfhtls\ (T0007 Release), 
and in the \bandj, \bandh , and \bandk\ bands of the near-infrared (NIR) from
the WIRcam Deep Survey \cite[\wirds,][Release T0002]{bielby12}.  
The resulting catalogue containing 11\,237 entries was carefully
cross-matched with ancillary X-ray through far-infrared (FIR) data, including \galex\ 
FUV/NUV\footnote{\tt http://www.galex.caltech.edu}
data from the four channels of \spitzer/\irac, FIR data from \spitzer/\mips\ at 24$\,\mu$m, 
and from PEP/\herschel\ \citep{lutz11} at 100 and 160 $\,\mu$m. Finally, photometric redshift (photo-z or \zp) 
and extinction estimates obtained from all these data using the {\it LePhare} code \citep{arnouts99,ilbert06},
as well as redshift data from other surveys (\deep\ and \cfhtls), were included in the dubbed \otelo\ multi-wavelength 
catalogue. Further details about this catalogue and other survey products are described in \oteloone.
For a quick reference, the main features of the survey and its catalogue are summarised in Table \ref{otelo_summary}.

\begin{table}[t]
\vspace*{2mm}
\caption[OTELO summary]{Main features of the \otelo\ survey and its multi-wavelength catalogue.}    
\vspace*{-5mm}
\label{otelo_summary}     
\begin{center}
\addtolength{\tabcolsep}{-1pt}
\begin{tabular}{l r}         
\hline   \\                      
 Parameter & Value   \\
\\[-0.5em]
\hline 
\\[-0.5em]
Field coordinates (RA, Dec.) & 14 17 33, +52 28 22\\
\\[-0.8em]
Effective surveyed area & 7.5\arcmin $\times$ 7.4\arcmin  \\
\\[-0.8em]
Wavelength range of RTF tunings & 9070 - 9280 ${\rm \AA}$ \\
\\[-0.8em]
RTF slice width ($\delta\lambda\,_{\rm FWHM}$) & 12 ${\rm \AA}$ \\
\\[-0.8em]
RTF sampling interval &  $\delta\lambda\;_{\rm FWHM}/2=6$ ${\rm \AA}$ \\
\\[-0.8em]
RTF slice effective passband ($\delta\lambda\,_{\rm e}$) & $\frac{\pi}{2}\ \delta\lambda\,_{\rm FWHM} \simeq$ 19.4 ${\rm \AA}$ \\
\\[-0.8em]
Wavelength accuracy & 1 ${\rm \AA}$  \\
\\[-0.8em]
Number of RTF tunings & 36 \\
\\[-0.8em]
Spectral range of pseudo-spectra & 230 \AA \\
\\[-0.8em]
Total integration time per slice & 6 $\times$ 1100 s \\
\\[-0.8em]
Mean PSF in RTF images & 0.87$\pm$0.13\arcsec \\
\\[-0.8em]
$m_{\rm lim}$ - \oteloint\ (3$\sigma$; 50\% compl.) & 26.38 mag.  \\
\\[-0.8em]
Mean error at $m_{\rm lim}$ & 0.33 mag.  \\
\\[-0.8em]
Total entries in the catalogue & 11237  \\
\\[-0.8em]
Number of ancillary data bands & 18 \\
\\[-0.8em]
Photo-z (\zp) accuracy & $\leqslant$ 0.2 (1+\zp)  \\
\\[-0.8em]
Astrometric accuracy (RTF images)&  0.23\arcsec \\
\\[-0.5em]
\hline                                           
\end{tabular}
\end{center}
\vspace*{-5mm}
\end{table}

Each source in the catalogue of \otelo\ is paired with exactly one pseudo-spectrum, which is defined 
as a vector composed of a succession of error-weighted averages of the fluxes measured on the individual 
RTF images at the
same wavelength in the tomography described above. Hence, a pseudo-spectrum of \otelo\ is formed by a
set of at least 36 flux elements evenly spaced by 6 \AA. Unlike spectra obtained by diffraction devices,
a pseudo-spectrum is the result of the convolution in the wavelength space of the input spectral energy
distribution (SED) of a given source, with the RTF instrumental response characterised by a succession
of Airy profiles. The wavelength sampling of \otelo\ pseudo-spectra is about one-third of
the effective passband $\delta\lambda\,_{\rm e}$ of an RTF slice (see Table \ref{otelo_summary}). Consequently,
line profiles in the pseudo-spectra are particularly affected by these instrumental effects and their flux values are 
more correlated in the wavelength space than standard spectra. In Section \ref{sec:deconvolve} we describe the 
methodology used to correct the pseudo-spectra for these effects. Figure 1 shows an\ example of a synthetic pseudo-spectrum 
with emission lines, as seen by \otelo\ at the optical centre of the RTF (see Section 2.1 in \oteloone).

\normalsize
\begin{figure}[h]
\centering
\includegraphics[angle=-90,width=\linewidth]{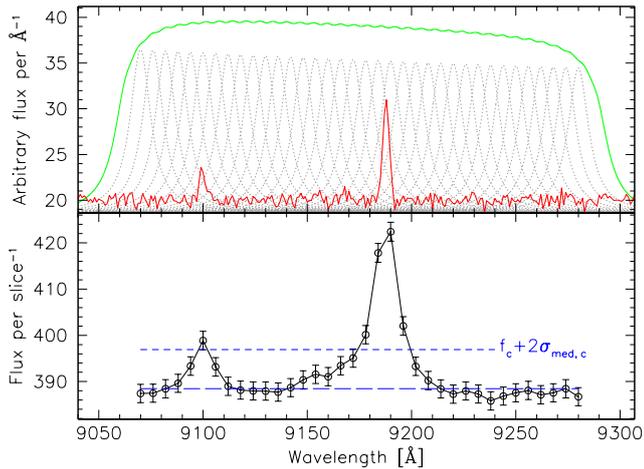}
\caption{Example of a synthetic pseudo-spectrum of an \oiii\ source as seen by \otelo. The upper panel
shows a synthetic spectrum (in red) with arbitrary flux density consisting of a flat continuum plus two 
emission lines with Gaussian profiles (FWHM = 2 \AA)
modelling the \oiiis\ doublet at z = 0.835, with Poissonian noise added. 
The grey dotted curves represent the Airy instrumental profiles of the \otelo\ RTF scan. These individual profiles 
can be synthesised into a custom filter response (green curve), which is useful to obtain an integrated IB flux. 
The lower panel contains the pseudo-spectrum obtained from the convolution of the synthetic spectrum in the 
upper panel with these Airy profiles, mimicking the \otelo\ data. The blue short-dashed line represents the 
2$\sigma_{\rm med,\,c\,}$ level above the median flux or pseudo-continuum (blue long-dashed line) defined in the main text.
Flux density between both spectra differs by a factor equal to the effective passband 
$\delta\lambda\,_{\rm e}$ of an RTF slice}.
\label{PSexample}
\end{figure}

\section{Selection and analysis of emission line sources}
\label{sec:ELSobs}

For the purposes of this paper, the following sections detail the steps required 
for selection and analysis of \otelo\ ELSs for any cosmic volume, based on the data products described above.

\subsection{Preliminary selection of emission-line sources in \otelo}
\label{preliminaryELS}

A given source in the \otelo\ multi-wavelength catalogue qualifies as a preliminary 
ELS candidate if (i) at least one point of the pseudo-spectrum lies above a value defined by f$_{\rm c}$ + 2$\sigma_{\rm med,\,c\,}$,
where f$_{\rm c\,}$ is the flux of the pseudo-continuum, which itself is defined as the median flux of the pseudo-spectra, 
and $\sigma_{\rm med,\,c\,}$ is the root of the averaged square deviation of the entire pseudo-spectrum with
respect to f$_{\rm c\,}$; and (ii) there is an adjacent point with a flux density above f$_{\rm c\,}$ + $\sigma_{\rm med,\,c\,}$.
 
Using these criteria, we found that  from a blind flux excess analysis of the pseudo-spectra a subset of 
6649 candidates match these conditions. This set contains all sources flagged as {\tt els\_preliminary} 
in the OTELO multi-wavelength catalogue.
The selection criteria given above set a limit on the performance of the 
algorithms used for retrieving line fluxes, whether via inverse deconvolution (see Section \ref{PSlinefit}) or automatic profile
fitting in pseudo-spectra as done by \cite{sanchezportal15}. 
Using a broad set of simulated \otelo\ pseudo-spectra such as those described in Section \ref{sec:PSlim}, we 
verified that if the intensity of the emission feature in the pseudo-spectra is below the thresholds indicated 
above, then none of these algorithms yield a plausible result.

However, there is a fraction of real ELSs whose emission line is truncated by the
limits of the \otelo\ spectral range but that do not meet the criteria cited above. According to the
RTF sampling and the wavelength range of the \otelo\ tomography, 
this fraction is about 5\% if the chance of the emission-line defined above is uniformly
distributed within the spectral range. These sources (as illustrated in Appendix \ref{PSexamples}), can also be 
retrieved using a (\bandz-\oteloint) colour--magnitude diagram, where \oteloint\ is
the integrated magnitude of a given source detected in the \otelodeep\ image (see Section \ref{sec:otelodesc}), 
as long as the contrast of the emission line above the pseudo-continuum
is large enough to give a reliable colour excess \cite[see][]{pascual07}. 

\subsection{Selecting emission line sources according to the emission line}
\label{PSlinefit}

As a consequence of the characteristic phase effect of the RTF, the passband of a given \otelo\
pseudo-spectrum is blueshifted (and slightly narrowed) as the corresponding source is further away
from the optical centre (see Eqs. 7 and 8 in \oteloone). Accordingly, the wavelength of the peak
flux $\lambda_{\rm\, peak}$ of an emission line in any \otelo\
pseudo-spectrum, independently of the mean distance of the source from the optical centre, can be
found between 9280 and 8985 \AA. The first of these latter two values is the reddest tuned wavelength of the \otelo\ scan
and the second is the phase-corrected wavelength corresponding to the bluest tuning at $r_{\rm max}$ = 4.2 
arcminutes from the optical centre, where $r_{\rm max}$ is the maximum angular distance from the optical
centre at which a source could be found in the \otelo\ field. Therefore, depending on the rest-frame 
wavelength of the emission line of interest, the 
wavelength range thus-defined allows us to determine the redshift window in which all the corresponding 
ELSs would fall.

Reliable redshift estimates are important for tagging the emission line(s) observed in \otelo\ 
pseudo-spectra, as is the case in any narrow band survey for the search of ELSs. The redshift 
of a source whose emission line falls inside the wavelength range defined above is a priori unknown. 
This can be solved using photo-z estimations, which nevertheless entails uncertainties. However, 
the by-products of these estimations also provide clues about the spectral classification of the sources 
to be studied. 

The multi-wavelength catalogue of \otelo\ provides, among others, photo-z estimations obtained by a $\chi^2$--SED 
fitting of the broad band ancillary data cited in Section \ref{sec:otelodesc}, both including and excluding 
the \otelodeep\ photometry. In both cases, the \zp\ uncertainty of each object was defined
by $\delta\,({\scalebox{.9}{\zp}}) = \lvert {\scalebox{.9}{\zp({\rm max})}} - {\scalebox{.9}{\zp({\rm min})}} \rvert / 2$,
where the maximum and minimum values of \zp\ correspond to the limits of a 68\% confidence interval of 
the \zp\ probability distribution function (PDF).

For selecting ELSs it is wise to employ the photo-z solutions that include the \otelodeep\ photometry (labeled {\tt z\_BEST\_deepY} in the catalogue) and its uncertainties. The overall accuracy of these photo-z estimations is 
better than $0.2\,(1+{\scalebox{.9}{\zp}})$ -- see Table \ref{otelo_summary} -- and the 1$\sigma$ 
width for ${\scalebox{.9}{\zp}} \lesssim 1.5$ is around
0.06. The redshift window defined above should be slightly broadened according to this width. All the sources 
with a photo-z value within the resulting redshift window, and that meet the condition given by 
$\delta\,({\scalebox{.9}{\zp}}) \leq 0.2\,(1+{\scalebox{.9}{\zp}})$, then pass to the next analysis stage.

It should be mentioned that the observed SED of a small but noticeable fraction of ELSs 
could be better fitted with AGN templates, as expected. The necessary data to compare the goodness-of-fit
of the photo-z depending on the template used are also given in the multiwavelength catalogue. For these cases, 
using a \zp\ obtained from the templates labeled {\tt AGN/QSOs} in \oteloone\ can help to complete the source selection 
according to the scientific goals pursued.

Once the pseudo-spectra of the preliminary ELS candidates in the corresponding photo-z range have been
identified, they must be further examined in order to (i) discard
artefacts or spurious sources (i.e.\ multiple peaks in pseudo-spectra linked to correlated sky noise, badly 
deblended sources in the \otelodeep\ image, objects near the boundary of the gap between detectors creating
false features in pseudo-spectra, and residuals of diffraction spikes from bright sources), and interlopers
with real emission lines (see Section \ref{sec:deconvolve}); (ii) flag other 
interesting spectral features or anomalies (i.e.\ line truncation or possible absorption lines); and 
(iii) assign a redshift value to the peak of the observed features in the pseudo-spectrum, or `guess' this value. 
The wavelength accuracy of this procedure is $\pm 3$ \AA, which translates to a redshift accuracy
of about 10\,$^{-3}$\,(1+z),  which itself is refined using
the inverse deconvolution algorithm described in Section \ref{sec:deconvolve}. 

A web-based graphic user interface (GUI) facility, which was prepared for the public release of \otelo\ data, can be used 
to carry out these tasks. This tool presents a complete 
dossier of every source in the survey, including observed SEDs, best-template fittings, broad-band image cutouts, 
cross-references with other public databases, and the pseudo-spectrum layer over an interactive line-identification tool. 
This tool was designed, among other applications, to obtain the `guess' redshift value once
the emission features are identified.  These steps prepare the ELS sample of interest, from which the final sample 
may be obtained.
 
\subsection{Deconvolving pseudo-spectra of emission-line sources}
\label{sec:deconvolve}

After the ELS sample of interest has been established, the pseudo-spectra of these ELSs must be
reduced by the instrumental response, using the guess redshift as input, in order to obtain useful data 
for science exploitation.

The \otelo\ RTF instrumental transmission is well described by Eq. 5 in \oteloone, and is represented in 
Figure \ref{PSexample}. As mentioned above, the effective passband ($\delta\lambda\,_{\rm e}$) adopted 
for the RTF survey is about three times the scan sampling. Thus, each point of an \otelo\ pseudo-spectrum is the 
result of a particular convolution of the input SED in the defined spectral range of the survey. 

Instead of a direct deconvolution of pseudo-spectra by the scan response function
(with the possible divergence and accuracy problems inherent to these techniques), we search for the
theoretical pseudo-spectrum that  minimises the error-weighted difference with the observed one. This 
inversion of spectral profiles has been successfully tested by other authors in similar situations \citep[e.g.][in the case
of the \sii\ doublet emission of H\,II regions in large design galaxies]{cedres13}.
In the case of this survey, this can achieved by combining a flat, zero-slope continuum with a set of three-parameter 
Gaussian profiles (central wavelength, amplitude, and width) for each emission line previously identified in the 
pseudo-spectrum (Section \ref{PSlinefit}), and then convolving this theoretical model with the instrumental response
of \otelo.
 
The line width and central wavelength of a theoretical Gaussian profile (one for each emission line identified)  
vary in broad ranges that are 
conveniently binned to obtain measurement resolutions of 0.25 \AA\ and 10\,$^{-4}$\,(1+z), respectively. This
redshift sampling provides an accuracy refinement of the guess value by a factor of approximately two after deconvolution, which is 
enough for the purposes of this work. The amplitude of these Gaussian profiles fluctuates in proportion to the flux error 
of the corresponding line peak in the pseudo-spectrum to be deconvolved. The initial value of the flux density of the 
continuum and the range within which  this value varies in the synthetic spectra 
are obtained starting from the f$_{\rm c}$ and $\sigma_{\rm med,\,c\,}$ values, respectively, which are defined
in Section \ref{sec:ELSobs}.
 
For a given observed pseudo-spectrum, a custom software application was used to perform the inverse 
deconvolution in the grid nodes of the above-defined parameter space. The theoretical spectrum associated with 
each ${\rm j}$-th node is convolved by the \otelo\ scan response, and the resulting pseudo-spectrum 
is compared with the observed one by means of the standard $\chi^2$ statistics, defined by

\begin {equation}
\label{eq:chisq}
\chi_{\rm j}^2=\sum_{\rm i=1}^{\rm N}\frac{(s_{\rm \,i} - t_{\rm \,i})^2}{\sigma_{\rm \,i}^2},
\end{equation}

\noindent where the sum is over the ${\rm N}$ slices of the pseudo-spectrum, $s_{\rm \,i}$ and 
$t_{\rm \,i}$ are the observed and theoretical flux densities, respectively, and $\sigma_{\rm \,i}$ 
is the observed flux density error at the ${\rm i}$-th slice. We adopted the ${\rm j}$-th model parameter set 
with the minimum $\chi^2_{\rm j}$ as the formal result of the analysis. 

We say that the deconvolution algorithm converges if a reliable theoretical spectrum is obtained from it, that is,  if (i) the errors 
on the line fluxes obtained from the inverse deconvolution are below 50\% of the line flux values, and (ii) the best theoretical 
continuum and EW of the line, or lines, are within the limits of the corresponding parameters obtained from the simulations 
described in Section \ref{sec:PSlim}. If both conditions are met, the ELS candidate becomes part of the final sample.

The parameter set given by the inverse deconvolution of the ELS candidates provides the 
best central wavelength of each emission line previously identified in the pseudo-spectrum and consequently a redshift 
estimation, \zotelo. This procedure also provides the best zero-slope continuum flux density and the parameters of the 
Gaussian function associated with each emission feature present in the pseudo-spectrum, allowing us in addition to 
determine the ${\rm m}$ ($<{\rm N}$) points of the pseudo-spectrum that contribute to each emission line. Using these data, 
the software application finally computes the net line flux by integrating the Gaussian profile 
obtained after $\chi^2$ minimisation because each profile is independent of the continuum. 
The error of the line flux is the average $\chi^2_j$ as defined by Equation \ref{eq:chisq}, but obtained over the
points of the pseudo-spectrum that contribute to the particular emission line. This approach is valid as long as there are no 
blended emission lines involved, which is the case of this work.

Finally, the inverse deconvolution of the pseudo-spectra allows us to recover a line flux measure -- even for the case of emission 
lines truncated 
by the spectral limits of the survey -- as long as the peak of the emission feature is within this range. An example of this situation 
is shown in Fig. \ref {PSexamples}. 


\section{Emission-line-source detection in \otelo\ from simulations}
\label{sec:PSlim}

Source detection in NB emission-line surveys depends not only on the emission line flux (determined in
the simplest case by the line width and intensity), but also on its strength relative to the continuum
parametrised by the observed equivalent width (EW$_{\rm obs}$).

We created synthetic spectra containing emission lines with Gaussian profiles distributed in a uniformly
gridded parameter space driven by the width and the intensity (amplitude or core) of the line, as well as the
flux density at the continuum. We then obtained simulated pseudo-spectra from the convolution of these
synthetic spectra with the instrumental transmission of \otelo. These simulations aim to determine educated
detection probabilities of ELSs for this survey. The limits of this parameter space are wide enough to
contain the preliminary ELS set described in \oteloone\ after the convolution of all the synthetic
spectra. Thus, the width of the lines (in terms of FWHM) was varied between 1 and 100 \AA\ in steps of
$\sim$4 \AA\ (200 km s$^{-1}$). The lower bound
corresponds to a width of $\sim$30 km s$^{-1}$ at the central wavelength of the RTF scan,
and the upper bound to a very broad line of $\sim$3300 km s$^{-1}$ at the rest-frame. The given range is
representative of the likely cases from dwarf galaxies to some broad-line AGNs. Similarly, the flux
densities $S_{\lambda}$ [erg s$^{-1}$ cm$^{-2}$ \AA$^{-1}$] of both the line intensity and the continuum of
the synthetic spectra were uniformly binned in the range $-22.3 \leq \log S_{\lambda} \leq -16.3$ in order to
widely sample the corresponding ranges of the preliminary ELSs of the survey.

Combined sky and photon noise per pixel was measured in 14 selected blank regions of 11 $\times$ 11 pixels
whithin the background noise maps resulting from the analysis of the 36 \otelo\ RTF slices. The noise 
values for each slice were averaged, obtaining a spectrum of the \otelo\
tomography. This noise spectrum was resampled to the resolution of the synthetic spectra. Each element of
the resulting noise spectrum was randomly modelled by a Gaussian distribution that approximates the
noise behaviour, and was then added to the corresponding wavelength value in the synthetic spectrum.

\begin{figure*}[t]
\centering
  \begin{tabular}{cc}
   \includegraphics[angle=0,width=0.48\textwidth]{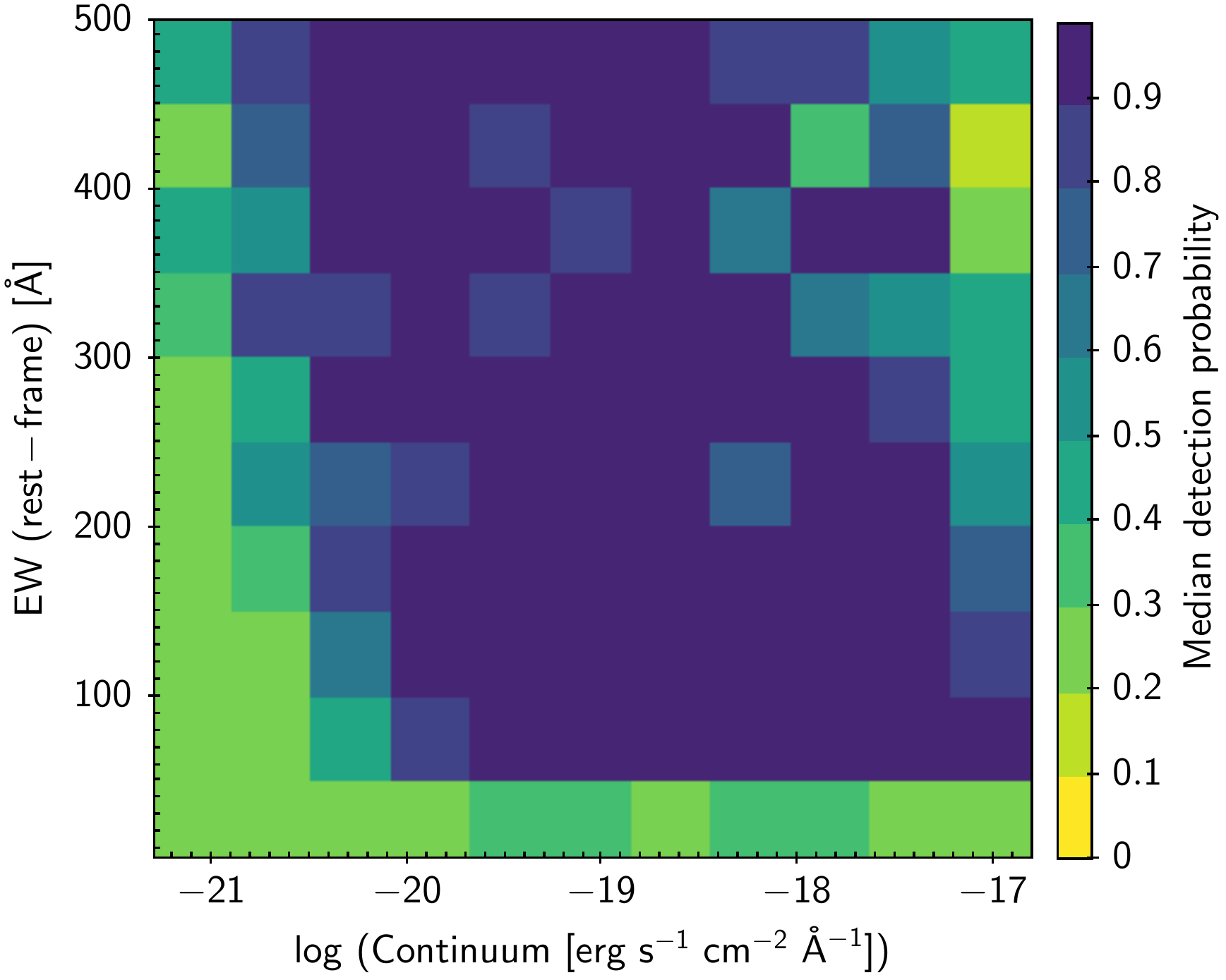} &
   \includegraphics[angle=0,width=0.48\textwidth]{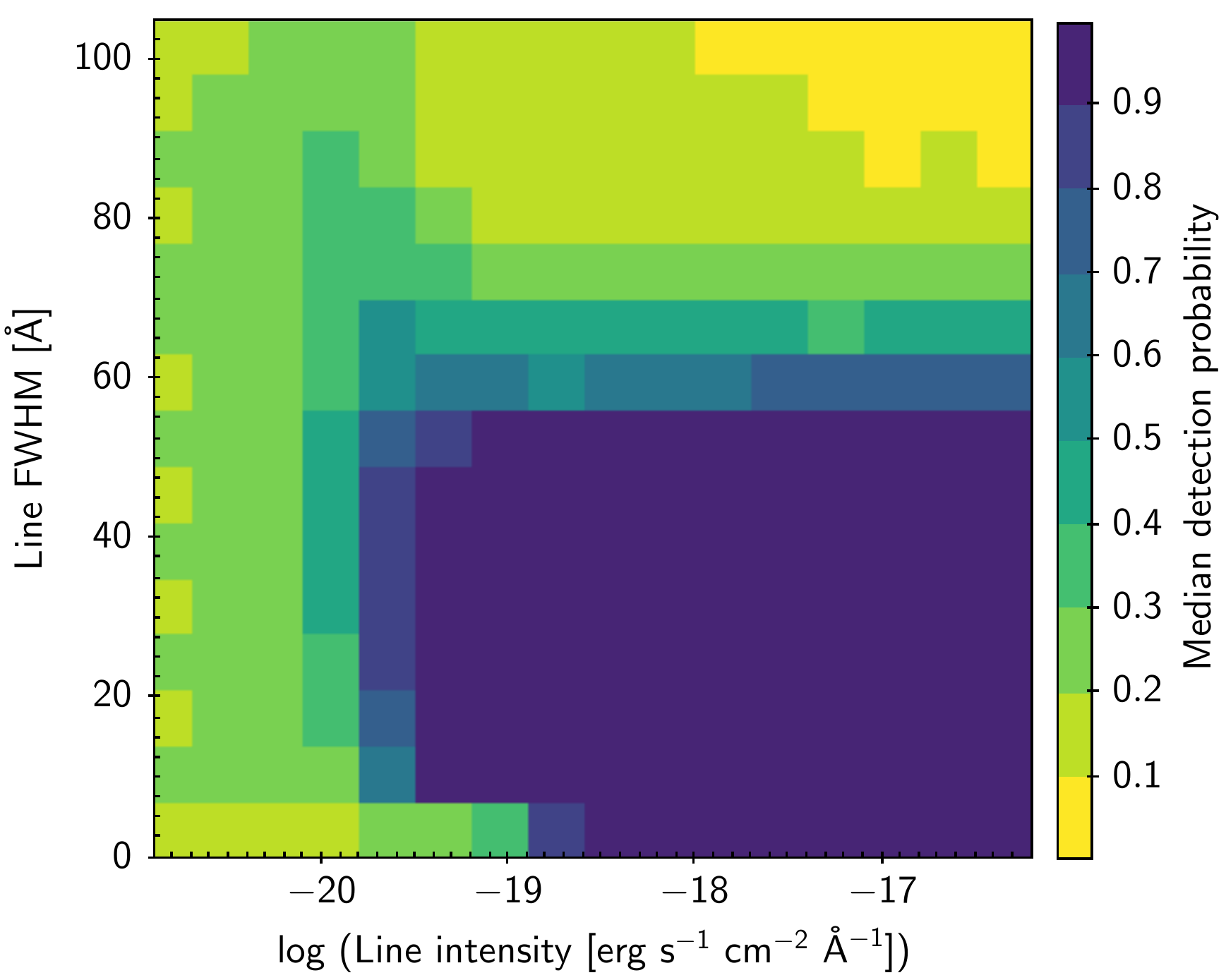} \\
  \end{tabular}
\caption{Examples of detection probabilities of ELSs in OTELO from simulations. Left: Median 
detected EW (rest-frame) as
a function of the continuum in the input spectra resulting from simulations.
Right: Median value of detection probability
in the line-width--intensity plane after the 500 runs of the simulations described in Sect. \ref{sec:PSlim}. 
The steep downturn of the median detection probability occurs when the emission-line width
exceeds $\sim$60 \AA.}

\label{simulaex}
\end{figure*}

After defining the line-width--intensity--continuum grid, a total of 500 independent synthetic spectra
per grid node  were obtained. Each synthetic spectrum was convolved with the instrumental response
of the RTF scan defined in Section \ref{sec:otelodesc} in order to obtain the simulated pseudo-spectrum.
For a given node of the grid we tested whether or not each of the 500 pseudo-spectra fulfilled the selection
criteria given in Section \ref{sec:ELSobs}. We assigned a value of `1' (or `0') whenever the line detection
was made (or not) and averaged the results in each node. This mean value defines the detection
probability function of the grid node in question, which depends on the line intensity, width, continuum
flux density, and the rest-frame EW calculated from these parameters.
As an example of the results, Figure \ref{simulaex} (left) shows the detection probability
distribution from the simulations in the EW (rest-frame)--continuum plane.

Data from simulations are also useful for obtaining relatively realistic emission-line flux distributions. Figure
\ref{simulaex} (right) represents the detection probability distribution in the line-width--intensity
plane, which allows us to calculate the required line flux statistics in each node of the simulation. Remarkably, this
plot confirms that the ELS detection probability drops dramatically when the line width of a source
exceeds $\sim$60 \AA. On the other hand, simulated data constitute the basis of the completeness
correction in the LF calculation given in Section \ref{sec:oiiiLF}.


\section{The case of \oiiis\ ELSs at $\langle$z$\rangle$=0.83}
\label{sec:oiii}

As mentioned above, the main aim of this work is to demonstrate the potential
of \otelo\ by analysing a sample of \oiiis\ ELSs at intermediate redshift. To this end, we first
describe the implementation of the selection steps given in Section \ref{sec:ELSobs}
in this particular science case, followed by relevant notes about the limitations of this selection method.
Finally, the main properties of the final \oiiis\ ELS sample are given.

\subsection{Selecting the \oiiis\ sample}
\label{sec:oiiifirst}

According to the content of Section \ref{PSlinefit}, we first defined 0.77 $\leq$ \zp\ $\leq$ 0.89 as 
the redshift interval needed to safely include all \oiiis\ ELS candidates. This range already takes into 
account the phase effect and the effective passband ($\delta\lambda\,_{\rm e}$) of the RTF slices 
at both ends of the scan, as well as the effect of the global accuracy of the photometric redshift estimations. Using the photo-z solutions 
that include the \otelodeep\ photometry, we found 401 preliminary \oiiis\ ELS candidates within this 
\zp\ range that meet the photo-z error condition also given in Section \ref{PSlinefit}. To this 
set we added 140 preliminary ELS candidates whose observed SEDs were better fitted with {\tt AGN/QSOs} templates
than with the galaxy library (see \oteloone).

The pseudo-spectra of the 541 resulting \oiiis\ ELS candidates were examined individually 
in order to identify plausible emission lines and assign a guess redshift to the source. From these, 
a total of 300 ELS candidates attributable to different
emission lines were found. In particular, the pseudo-spectra of 208 of 
these ELS candidates contained at least one of the \oiii\ features, and the emission feature in the pseudo-spectra of 
92 candidates were assigned to neighbouring
emission lines (see Table \ref{oiiiselection}) based on the significance and the error estimation of the \zp\ 
and the information provided by the pseudo-spectra.
From the remaining 241 sources, 21 
were discarded as probably spurious and 220 could not be classified as reliable ELS candidates owing to the
presence of multiple unlikely features in their pseudo-spectra, in accordance with the requirements given in 
Section \ref{PSlinefit}. 

\begin{table}[ht]
\vspace*{2mm}
\caption[OIII ELS selection]{Number of ELS candidates obtained from the selection process of the \oiiis\
emitters.
}
\vspace*{-5mm}
\label{oiiiselection}
\begin{center}
\addtolength{\tabcolsep}{-3.5pt}
\begin{tabular}{l c}
\hline   \\
ELS candidate subset\, & \,Number \\
\\[-0.5em]
\hline 
\\[-0.5em]
\oiiis\       & 208 \\
\\[-0.7em]
\hb\       & 56 \\
\\[-0.7em]
\heiihb\       & 25 \\
\\[-0.7em]
\ni\       & 6 \\
\\[-0.7em]
\oiiia\       & 5 \\
\\[-0.7em]
\hline
\\[-0.7em]
Total & 300 \\
\\[-0.7em]
\hline
\end{tabular}
\end{center}
\end{table}
\normalsize

The pseudo-spectra of the 208 \oiiis\ ELS candidates were deconvolved in order to obtain accurate
redshifts (\zotelo), continuum flux densities, and parameters for the emission lines(s) detected.
The algorithm successfully converged for a total of 182 sources.

We make a comparison between the resulting redshifts from \otelo\ and those obtained from
\deep\ data. Catalogued \deep\ redshifts with a quality flag Q $\geq$ 3 \citep{newman13} are also 
available for a total of 26 of these latter-mentioned 182 sources. The Pearson correlation coefficient 
between spectroscopic redshifts from \otelo\ and \deep\ surveys is 0.998, with no outliers. The mean accuracy 
of \zotelo\ related to \zs, $\vert\Delta {\rm z}\vert/(1+{\rm z}_{\rm\, DEEP2}) \leq 2.54\, \times\, 10^{-4}$. 
This value is consistent with the upper-limit error of ${\rm z}_{\rm\, DEEP2}$ with Q $\geq$ 3, which
is on the order of  3 $\times$ 10$^{-4}/(1+{\rm z}_{\rm\, DEEP2})$.

Representative examples of \oiiis\ ELSs with a bright, faint, or very faint continuum, and their deconvolved
pseudo-spectra, are shown in Figure \ref{PSexamples}. As also shown in this figure, the inverse deconvolution 
provides recovery of the emission line profile truncated by the edges of the spectral range of \otelo\
whenever the emission line peak was sampled in the pseudo-spectrum.

\subsection{Contamination of the \oiiis\ sample}
\label{sec:oiiicontamination}

Galaxy count statistics and other empirical estimations of cosmological parameters depend on the results of
low-resolution emission-line galaxy (ELG) surveys that often contain foreground and background contaminants or interlopers due to
misidentifications of spectral features \citep{hayashi18, grasshorn19}. In particular, a detailed estimation of 
the residual contaminants possibly present in the final sample of ELSs used in this work is a necessary but not
sufficient condition for a reliable assessment of the LF(\oiiis).

The most common contaminants of NB surveys for the search of ELSs are the possible Balmer-jump sources,
as well as the lower and higher redshift interlopers \citep[see e.g.][]{hayashi18} if the errors in 
photo-z are comparable to the differences between these redshifts and the one being sought. In the science 
case presented here, this class of contaminants is composed of misidentified \ha\ and \oii\ ELSs at redshifts of 
$\sim$0.4 and $\sim$1.4, respectively. We also envisage, as an additional class of interlopers in \otelo, the 
ELS preliminary candidates selected on the basis of inadvertently incorrect photo-z estimations, despite  their 
error values fulfilling the criteria given in Section \ref{PSlinefit}. As mentioned above, the selection of ELSs in \otelo\ 
depends on the photo-z and the associated errors, 
as well as the identification of emission lines in pseudo-spectra above a given significance level.

The sample of \oiii\ ELSs previously obtained is free from Balmer-jump interlopers because the latter ones can easily 
be identified among the preliminary ELS candidates in the case where a misleading photo-z (see below) has 
allowed them to be included in the sample. Therefore, the degree of contamination of the final \oiiis\ ELS sample 
is limited only to determine the balance of interlopers due to known line misidentifications and the likely fraction 
of sources with catastrophic photo-z (and reasonable errors) incorrectly assumed as \oiiis\ ELS.

In order to reject redshift interlopers, we selected all the preliminary \ha\ and \oii\ ELS candidates in \otelo\  whose 
photo-z estimations lie within 0.36 $\leq$ \zp\ $\leq$ 0.42 and 1.41 $\leq$ \zp\ $\leq$ 1.55, respectively. These sets
contain the broadest selection of possible redshift interlopers present in the \oiii\ ELS sample obtained by following
the steps described in Section \ref{sec:oiiifirst}. We then cross-correlated the \oiii\ ELS sample with each one of
these sets. We found no \ha\ candidates in the \oiii\ sample from the diagnostics of the lower redshift interlopers.
On the other hand, we found two possible \oii\ interlopers (i.e. the \otelo\ sources {\tt ID}: 03564 and 04264) after 
cross-correlation of the \oiii\ ELS sample with the preliminary \oii\ ELS set. However, the source {\tt ID}: 03564 
is a \oiiib\ + \hb\ ELS, with ${\rm z}_{\rm\, DEEP2}=0.8348$, and the source {\tt ID}: 04264 clearly exhibits 
the \oiii\ doublet in its pseudo-spectrum. Thus, the photo-z estimations of these sources differ from the true redshift. 
In both cases, high-resolution images from \acs\ reveal the presence of sources with several knots inside the \otelo\ 
segmentation imprint. This fact suggests that the observed SED of such sources is of composite type and may not be
necessarily well adjusted by the single galaxy templates used to estimate their photo-z. Regarding the results of this 
analysis, there are not indications of lower or higher redshift interlopers in the \oiii\ sample obtained as described in 
Section \ref{sec:oiii}.

Finally, the adoption of the uncertainty limits on photo-z given in Section \ref{PSlinefit} ensures that ELS candidates 
with large errors are excluded from the final sample. However, there is a non-zero probability that some ELS candidates 
have been selected on the basis of catastrophic redshifts despite their moderated errors. Hence, and according to 
the established in \oteloone, the fraction of these likely interlopers ($\delta_{\rm cat-z}$) up to 
z$\sim$1.4 is at most 0.04. This statistical contamination effect is included in the error budget of 
the LF(\oiiis) formulated in Section \ref{sec:oiiiLF}.  

\subsection{The \oiiis\ doublet and the missing emission-line sources}
\label{sec:oiiieye}

Depending on the signal-to-noise ratio of the emission lines, an \oiii\ source can be easily recognised 
in \otelo\ pseudo-spectra without additional information if both components of the doublet appear in it, because 
the characteristic ratio \oiiib/\oiiir\ $\approx$ 0.3 \citep{osterbrock06}. This distinctive feature can help 
to quantify the main bias of the ELS selection procedure used above in the practical case of \oiiis\ and
other chemical species.

Using the web-based GUI in a procedure completely separate from the \oiiis\ ELS selection, all the 
pseudo-spectra were visually inspected to look exclusively for the \oiiis\ doublet characteristic signature. We 
found 29 sources whose pseudo-spectra are fully compatible with the \oiii\ emission. All of these 
visually selected \oiiis\ ELSs 
are included in the 182 \oiiis\ obtained as described in Section \ref{sec:oiiifirst} except for two \oiiis\ emitters 
for which there are no detections in ancillary data because of their faintness at the continuum; this 
prevents any kind of \zp\ estimation, and consequently such sources cannot be selected by the 
procedure described in Section \ref{sec:ELSobs}. The inverse deconvolution of both of these latter two 
mentioned pseudo-spectra thoroughly 
converged, as expected, and these are shown in Appendix \ref{PSexamples}. These bona fide \oiii\ sources
were added to those from the standard selection process. In this way, the final set of\oiiis\ ELSs in \otelo\ 
was found to contain 184 sources.

According to the emission features observed in the pseudo-spectra, a total of 84 ELSs show the \oiii\
doublet in their pseudo-spectra. The remaining 100 ELSs to complete the final set of \oiiis\ ELSs in \otelo\ 
show either the \oiiir\ or the \oiiib\ emission line. For illustrative purposes, the \otelo\ pseudo-spectra are 
wide enough to contain the \oiiib\ and the \hb\ lines near their extremities at the mean redshift of this science case. 
Thus, among the ELSs that only show the \oiiib\ component in their pseudo-spectra (34), we account for 23 
ELSs that exhibit the \oiiib\ and \hb\ pair. The study of the \hb\ ELS sample in \otelo\ is the subject of a forthcoming 
contribution (Navarro et al., in preparation). Table \ref{subsetsoiii} summarises the statistics of the subsets 
that integrate the final sample of \oiiis\ ELSs, depending on the emission features found.

\begin{table}[ht]
\vspace*{2mm}
\caption[OIII ELS selection]{Number counts in subsets that contribute to the final \oiiis\ ELS sample of \otelo\  according to the features observed in their pseudo-spectra.}
\vspace*{-5mm}
\label{subsetsoiii}
\begin{center}
\addtolength{\tabcolsep}{0pt}
\begin{tabular}{l c}
\hline   \\
ELS subset & Number \\
\\[-0.5em]
\hline 
\\[-0.5em]
\oiii & 84 \\
\\[-1.0em]
(includes 2 ELSs missing in ancillary data) & \\
\\[-0.7em]
\oiiir\ only & 66 \\
\\[-0.7em]
\oiiib\ only & 11 \\
\\[-0.7em]
\oiiib\ + \hb & 23 \\
\\[-0.7em]
\hline \\[1 pt]
Total & 184 \\
\\[-0.7em]
\hline
\end{tabular}
\end{center}
\end{table}
\normalsize

Based on these number statistics on the \oiiis\ ELS detections, the probability $p_{\rm d}$ of 
missing ELSs that show the \oiiis\ doublet in \otelo\ is 2/84 or 2.4\% if the ELS selection procedure described in 
Section \ref{sec:oiiifirst} is followed without further add-ons. Under the same hypothesis, and considering
a uniform distribution of emission lines in \otelo\ pseudo-spectra at <z>=0.83, the probability 
$p_{\rm s}$ of missing ELSs whose pseudo-spectra show either the \oiiib\ or the \oiiir\ emission line is 
around 2\%. If these estimations are considered valid for any other emission line in
\otelo, the probability of excluding a true ELS is $p_{\rm doublet} + p_{\rm single}$ or about 4.4\%,
because both events are disjointed. As mentioned, in the case considered in this paper this fraction drops 
to 2\%, since we already added the two \oiiis\ ELSs identified exclusively from pseudo-spectra 
inspection.

At last, it is worth pointing out that the error-weighted mean flux ratio $f$(\oiiib)/$f$(\oiiir)
of the 84 sources whose pseudo-spectrum shows the doublet is 0.328, with a standard deviation
$\sigma$ = 0.229. This mean ratio is constant below 1$\sigma$ over the whole range of
EW$_{\rm obs}$(\oiiir), which is in good agreement with the theoretical ratio given above, despite the
dispersion of measured ratios increasing for lower EW$_{\rm obs}$. Similar behaviour is reported in
\cite{boselli14}, who attribute some of the greatest deviations to AGN hosts with low EW(\oiiir).
Hereafter, the \oiiis\ line fluxes refer to \oiiir, unless otherwise specified.

\subsection{Main physical properties of \oiiis\ emission line sources}
\label{sec:O3ELS}

From the previous analysis, the final \otelo\ sample of \oiiis\ ELSs is distributed in the redshift range 
0.78 $\leqslant$ \zotelo\ $\leqslant$ 0.87, with a mean $\langle {\rm z}\rangle$ = 0.83. 

\begin{figure*}[t]
\centering
  \begin{tabular}{cc}
   \includegraphics[angle=-90,width=0.48\textwidth]{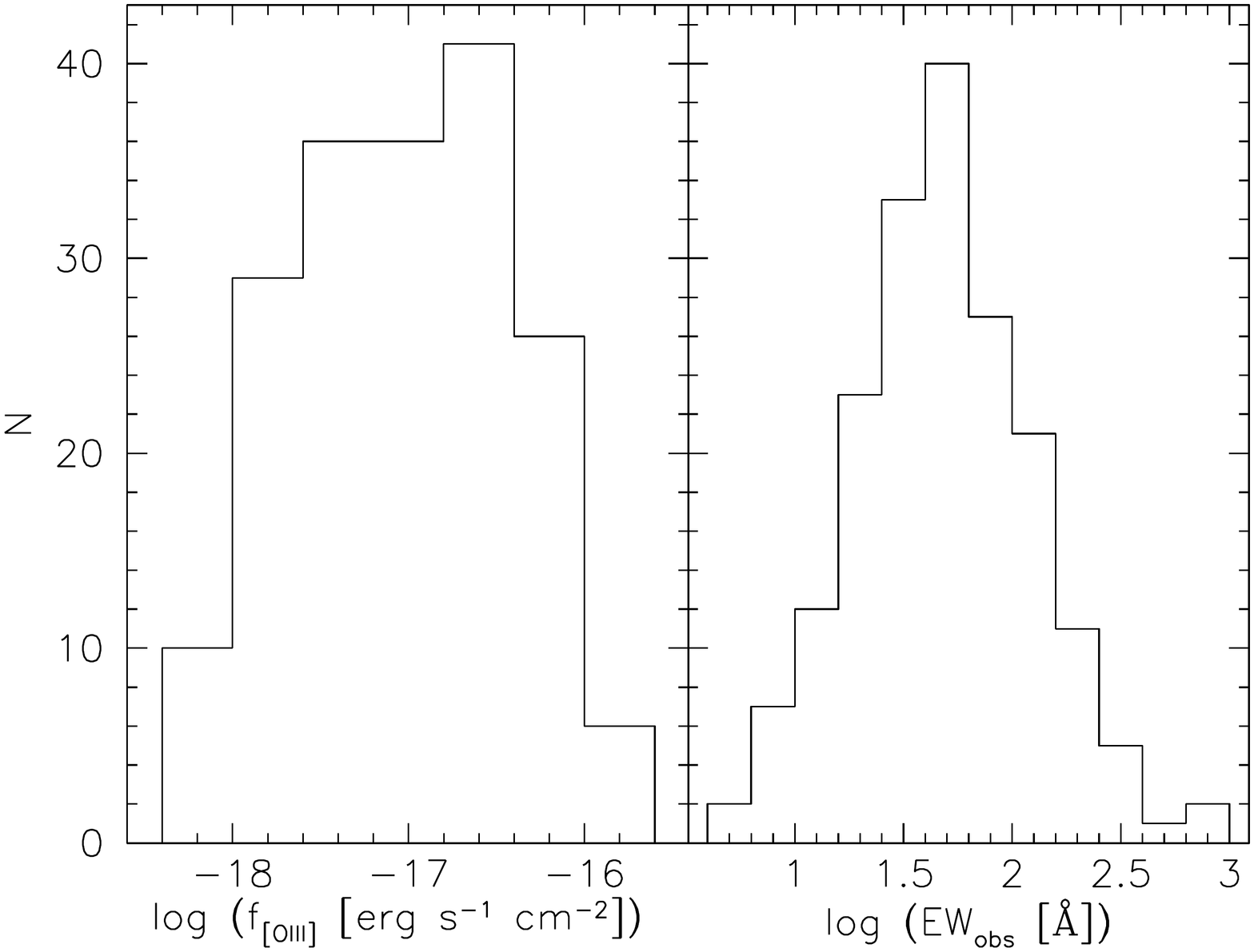} &
   \includegraphics[angle=-90,width=0.48\textwidth]{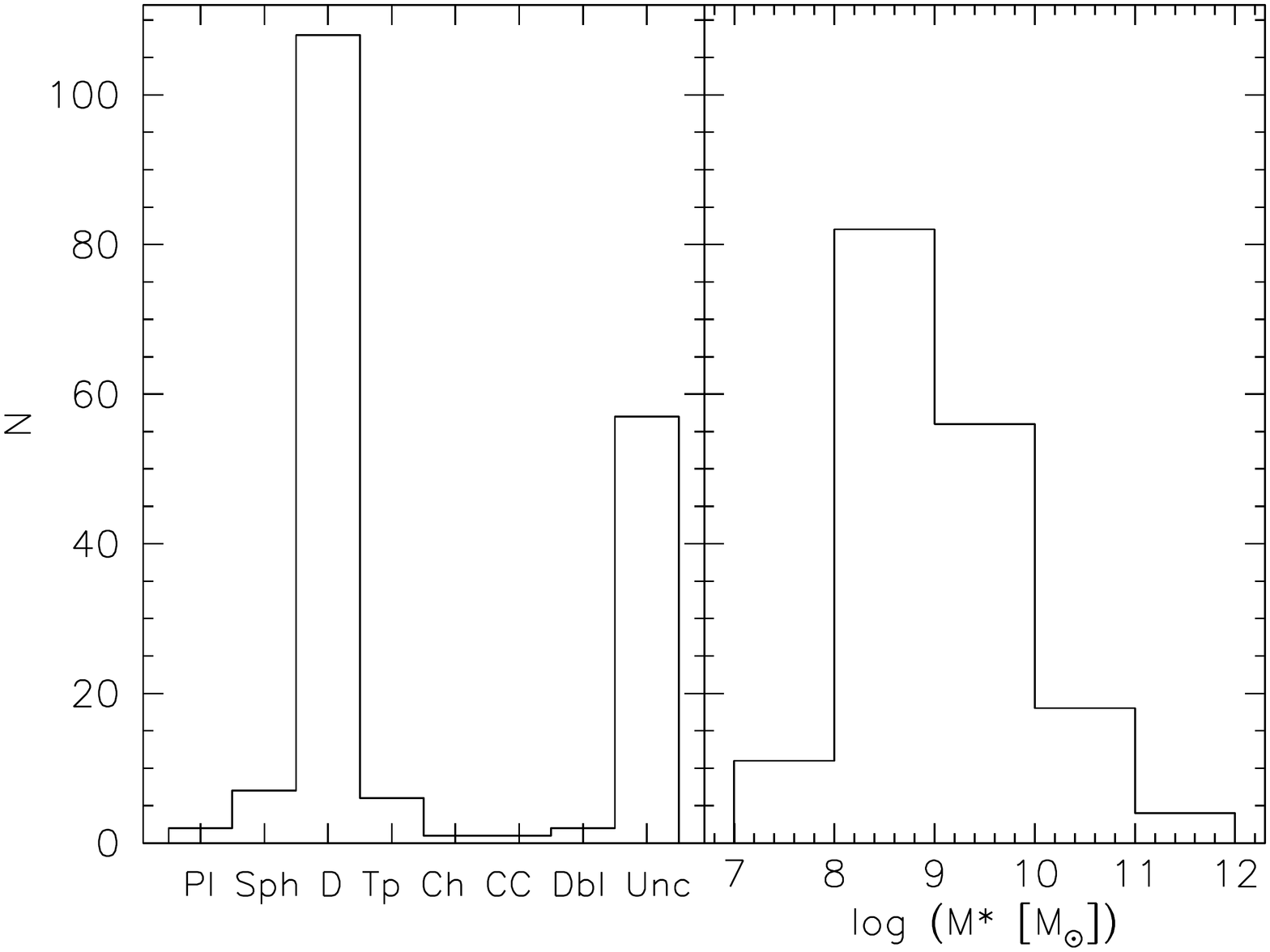} \\
  \end{tabular}
\caption{Distribution of characteristic parameters of the OTELO \oiiis\ sample. Left: \oiiis\ line flux and 
observed EW measured in \oiiis\ pseudo-spectra. Line flux and EW$_{\rm obs}$ limits are 
$\sim$5 $\times$ 10$^{-19}$ erg s$^{-1}$ cm$^{-2}$ and $\sim$6 \AA. 
Right: Morphology and stellar mass distribution of the OTELO \oiiis\ ELS sample.
The main morphological classes represented here are: \textsf{Pl} - Point-like; \textsf{Sph} - Spheroid; 
\textsf{D} - Disk; \textsf{Tp}- Tadpole; \textsf{Ch} - Chain; \textsf{CC} - Clumpy Cluster; 
\textsf{Db} - Doubles; \textsf{Unc} - Uncertain / Non-classified. At least 127 sources are
disc-like galaxies and the median stellar mass is 8.36 $\times$ 10$^{8}$ M$_{\odot}$. See main text for details.
}
\label{samplelimits}
\end{figure*}

Figure \ref{samplelimits} 
(left) shows the line flux and 
the EW$_{\rm obs}$ distributions of this set. The median values are 
8.7 $\times$ 10$^{-18}$ erg s$^{-1}$ cm$^{-2}$ and 45.7 \AA, 
respectively. The \oiiis\ ELS sample reaches a limiting line flux of 
4.6 $\times$ 10$^{-19}$ erg s$^{-1}$ cm$^{-2}$ and EW$_{\rm obs}$ as low as $\sim$5.7 \AA. These limits are 
consistent with those expected when derived from the simulations presented in Section \ref{sec:PSlim}, and both
could be taken as representative values of the survey.

Figure \ref{samplelimits} (right) shows the morphology and stellar mass distribution of the sources in 
the \oiii\ ELS set.  From a visual and quantitative morphological analysis of \otelo\ sources 
(Nadolny et al., in preparation) using publicly available high-resolution images from 
\acs814\footnote{\url{http://aegis.ucolick.org/mosaic_page.htm}}, 85\% of the 127 
classified sources (from the 184 ELSs of the final sample) are composed of disc-like galaxies. For
57 sources of this ELS sample the morphological classification was uncertain because of their exceedingly 
low signal level in the image, or were simply left unclassified because they are outside the \acs\ imprint.
The visual classification of all sources up to \zp\ = 2 is being performed by the \otelo\ Team and involve 
the use of MorphGUI, an application specially developed for this purpose by \cite{kartaltepe15}, 
in the framework of CANDELS. A list of defined morphological classes is given in 
Figure \ref{samplelimits} (right). 

We obtained stellar mass (${\rm M}_\star/{\rm M}_{\odot}$) estimates for 171 of 184 \oiii\ ELSs (94\%) using the relation of
mass-to-light ratio with colour for star-forming (SF) galaxies at redshift z $<$ 1.5 given in 
\cite{lopezsanjuan19}. Following this prescription we employed the rest-frame (\bandg-\bandi) colour, 
together with absolute \bandi-band magnitude for each source. The required data were computed from the 
available observed data in the \otelo\ multi-wavelength catalogue and the k-correction was provided
by a SED-fitting of these photometric data at the redshift obtained from inverse deconvolution or \zotelo. The 
reported dispersion of this mass-to-light ratio ($\sigma_{\rm SF}$ = 0.9) is slightly smaller than that obtained 
by \cite{taylor11} for galaxies 
from the GAMA survey, probably because of the addition of a quadratic colour term in such a relation. Therefore, based 
on this proxy, almost  87\% of the 171 \oiii\ ELSs should belong to the low-mass galaxy population 
(${\rm M}_\star \lesssim 10^{10}$ M$_{\odot}$). Furthermore, using the observational criterion given by \cite{gildepaz03} for
a galaxy to be classified as a blue compact dwarf (BCD; i.e.\ $M_{\rm K}>21$ mag), 74\% of the \oiiis\ ELS sample
is composed of dwarf galaxies.

Finally, 31 (17\%) of the  184 \oiiis\ ELSs are infrared (IR) galaxies, based on the \spitzer\
and \herschel\ data fit using \cite{chary01} IR SEDs (see \oteloone\ for additional details). Their
stellar masses are greater than 10$^{9}$ M$_{\odot}$ and 11 of them qualify as luminous IR galaxies
(LIRG: 10$^{11}$ $<$ $L$({\rm TIR} [$L_{\odot}$]) $<$ 10$^{12}$).


\section{The \oiiis\ luminosity function}
\label{sec:oiiiLF}

Using the data obtained from the analysis of the \oiiis\ ELS set described, as well as
suitable information from the simulations of the survey, we compute here the \oiii\ luminosity 
function, LF(\oiiis), of \otelo\ at $\langle$z$\rangle$=0.83 after taking into account the main
uncertainties and selection effects that could impact this calculation, which are described as follows.

\subsection{Completeness of the \oiiis\ sample}
\label{sec:completeness}

There are two main considerations to be made concerning the completeness of the final \oiiis\ ELS sample. 
Firstly, the probability of excluding a true positive \oiiis\ ELS because of its 
faintness at the continuum should be taken into account, 
as discussed in Section \ref{sec:oiiieye}; this fraction, $\delta_{\rm miss}$, accounts for 2\% of this 
particular ELS sample. Secondly, looking at the bounds imposed by data in the final sample, and based on 
emission-line object detectability simulations in the EW$_{\rm obs}$--continuum binned space described in Section 
\ref{sec:PSlim}, we obtained mean values of the ELS detection probability (MDP) as a function of the 
\oiiis\ line flux. This probability function was in turn modelled by a sigmoid algebraic function (similar in behaviour to  
the error function, {\tt erf}), of the form

\begin{figure}[ht]
\centering
\includegraphics[angle=-90,width=\linewidth]{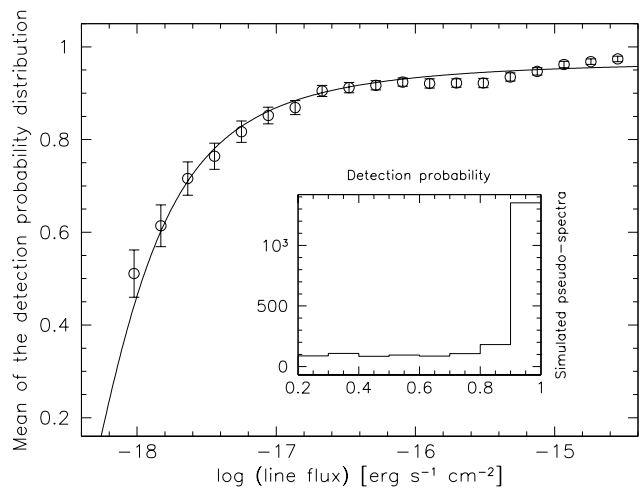}
\caption[]{Mean of the detection probability of the \oiiis\ ELSs at $\langle$z$\rangle$ = 0.83. The inset shows the
detection probability distribution of all simulated pseudo-spectra described in Sect. 4 used to obtain the function represented
here. The curve traces the
least-squares weighted fit of the sigmoid function used to model such flux-binned mean values.
Bars represent the mean standard error used for fit weighting.}
\label{detprob}
\end{figure}

\begin{center}
\begin{equation}
d=\frac{a F}{\sqrt{c+F^{2}}},
\label{completeness}
\end{equation}
\end{center}

\noindent where $F=\log(f_l)+b$, $f_l$ is the line flux, and $a=0.972\pm0.007$, $b=18.373\pm0.092$, and $c=0.475\pm0.122$ 
are the parameters obtained from a weighted least-squares fit. This function was adopted as the completeness
correction function (CCF) for calculations involving emission-line flux statistics at \zred\ = 0.83. The CCF
and its parent MDP distribution are shown in Figure \ref{detprob}. The mean standard error of this
distribution is $\delta_{\rm MDP}=0.017$.

\subsection{Cosmic variance}
\label{sec:cv}
As a deep pencil-beam ($<$1 deg$^2$), single-pointing survey, the \otelo\ scan is affected by
the relative cosmic variance (CV) at different levels, which can be ascribed to the effects of the 
underlying large-scale density fluctuations. In practice, this uncertainty depends on the mass (or luminosity) 
and the comoving volume sampled. Typically, CV is $\lesssim$10\% for a single volume 
of $10^{7}\ h_{0.7}^{-3}$ Mpc$^3$ projected onto  an almost square sky region \citep{driverrobot10}. In nearly 
all \otelo\ science cases, as  addressed in this paper, the volumes
explored are significantly smaller. Therefore, an estimate of the uncertainty due to the CV is imperative.

We estimated an uncertainty function due to the CV, $\sigma_{\rm CV}({\rm M}_\star)$ for eight bins in the
stellar mass range 7.5 $\leqslant$ $\log({{\rm M}_\star/{\rm M}_{\odot}})$ $\leqslant$ 11.5 for the \oiiis\ 
sample, obtained as
described in Section \ref{sec:O3ELS}, by following the prescription of \cite{moster11}. For practical
purposes, the uncertainties obtained per mass bin were fitted by a three-parameter power law.
To give a general idea of the CV effect, the mean uncertainty obtained for our ELS sample 
at $\langle$z$\rangle$=0.83 is $\langle$$\sigma_{\rm CV}$$\rangle$=0.396.

\subsection{Survey volume}
\label{sec:volume}

With regard to the phase effect mentioned in Section \ref{PSlinefit}, for each \oiiis\ ELS
there is a corresponding passband with limits and width in wavelength (and hence redshift) that depends on 
the position of the source with respect to the optical centre of the RTF. Thus, with a redshift range 
for each source as a function of the radial distance to the optical centre (e.g.\ such as the top-hat profile shown
in Fig. \ref{PSexample}), and taking into account the effective
passband $\delta\lambda\,_{\rm e}$
of the RTF in a single slice, we obtained the maximum comoving volume in which the emission line(s)
of such sources can be detected. The comoving volumes range from 6708.9 (for a hypothetical source
at the optical centre) to 6540.3 Mpc$^{3}$ (for a source located at $r_{\rm max}$). The comoving
volume at a radius that divides the field of view into two equal angular surfaces is 6633.4 Mpc$^{3}$,
and this is the value adopted as representative of the \oiiis\ survey volume in Table 
\ref{otelo_lf_values}. The effect of the wavelength accuracy of the RTF (see Table \ref{otelo_summary}) on comoving volume 
calculations is negligible.

\subsection{Luminosity function estimation}
\label{sec:oiiiLFcalc}

The \oiiis\ luminosity of each ELS of the sample is given by $L({\rm \oiiis}) = 4\, \pi\, {\rm f_{\oiiis}}\, D^2_{L}$,
where ${\rm f_{\oiiis}}$ is the \oiiis\ flux obtained from the inverse deconvolution of the pseudo-spectrum 
(i.e.\ with no correction for dust attenuation in the galaxy), and $D_L$ is the luminosity distance. The \oiiis\ 
luminosity was distributed in the range $39.2 \le \log L({\rm \oiiis})\, [{\rm erg\, s^{-1}}] \le 42.0$. 

The first task for the LF calculation is to compute the number $\Phi$ of galaxies per unit volume
($V$) and per unit \oiiis-luminosity $\log L$(\oiiis). In this case, this number is provided by 

\begin{center}
\begin{equation}
\label{otelophi}
\Phi[\log L({\rm \oiiis})] = \cfrac{4\pi}{\Omega}\; \Delta[\log L({\rm \oiiis})]^{-1}\; \sum\limits_{\rm i}^{} \cfrac{1}{V_{\rm \,i}\; d_{\rm \,i}},
\end{equation}
\end{center}

\noindent where $d_{\rm \,i}$ is the detection probability given by the CCF, $V_{\rm i}$ is the comoving volume 
for the $i$-th source, $\Omega$ is the 
surveyed solid angle ($4.7\, \times\, 10^{-6}$ str), and $\Delta[\log L({\rm \oiiis})] = 0.4$ 
is the adopted luminosity binning. 

Instead of using methods of measured data re-sampling, we adopt a scheme of error propagation based on the 
potential sources of uncertainty described in the preceding sections. The $L({\rm \oiiis})$ bins were treated
independently. Estimation of the total error per bin as given in Table \ref{otelo_lf_values} was done as 
follows. 
 
The main uncertainties that affect the LF(\oiiis) are those linked to the hypothesis of a Poisson point 
process in the galaxy distribution, which depends on the number of ELSs per $L({\rm \oiiis})$ bin and the 
effects of the CV ($\sigma_{\rm CV}$). The total error per bin also includes the contributions of (i) the mean 
standard error of the completeness correction ($\delta_{\rm MDP}$), (ii) the false-positive detection rate 
of \oiiis\ ELSs ($\delta_{\alpha-1}$), (iii) the probability of incorrect ELS identification ($\delta_{\rm cat-z}$), 
and (iv) the lost \oiiis\ ELS fraction due to a selection method based in part on broad-band photometric 
data ($\delta_{\rm miss}$). The contributions of these errors were summed in quadrature and they are given 
in Table \ref{otelo_lf_values}, along with the LF(\oiiis) values obtained from Equation \ref{otelophi}.

\begin{table}[h]
\vspace*{2mm}
\caption[Luminosity function values]{Binned values of the observed \oiiis\ luminosity function obtained from
Equation \ref{otelophi}. Uncertainties correspond to the total errors described in the text.
The last column contains the observed number (i.e.\ before completeness correction) of the \oiiis\ ELSs in each 
luminosity bin.
} 
\vspace*{-5mm}
\label{otelo_lf_values}
\begin{center}
\begin{tabular}{c c c}
\hline   \\
$\log L($\oiiis$)$  & $\log \phi$             & Number of \\
\\[-0.9em]
$ $[erg s$^{-1}$]      & [Mpc$^{-3}$ dex$^{-1}$] & \oiiis\ ELSs \\
\\[-0.5em]
\hline     \\[-0.7 pt]
 39.4 & -1.626$^{+0.151}_{-0.234}$ & 14 \\
\\[-0.5em]
 39.8 & -1.758$^{+0.136}_{-0.200}$ & 30 \\
\\[-0.5em]
 40.2 & -1.709$^{+0.133}_{-0.192}$ & 42 \\
\\[-0.5em]
 40.6 & -1.836$^{+0.136}_{-0.198}$ & 34 \\
\\[-0.5em]
 41.0 & -1.760$^{+0.134}_{-0.195}$ & 42 \\
\\[-0.5em]
 41.4 & -2.161$^{+0.150}_{-0.230}$ & 17 \\
\\[-0.5em]
 41.8 & -2.698$^{+0.194}_{-0.359}$ & 5 \\
\\[-0.7em]
   
\hline
\end{tabular}
\end{center}
\end{table}
\normalsize

The \cite{schechter76} function is the formalism adopted here to describe the luminosity function, 
$\phi(L)\, {\rm d}L = \Phi[\log L({\rm \oiiis})]\, {\rm d}(\log L)$, which is defined by

\begin{center}
\begin{equation}
\phi(L)\, {\rm d}L = \phi^\ast (L/L^\ast)^\alpha \exp(-L/L^\ast)\, {\rm d}(L/L^\ast).
\end{equation}
\end{center}

The parameter $L^\ast$ is the characteristic value that separates the high(exponential)- and low(power-law)- luminosity regimes in the LF driven by $\phi^\ast$ and $\alpha$, which are the 
number density at $L^\ast$ and slope of the faint end of the function, respectively. This
parametrisation facilitates the calculation of the luminosity and number density of the galaxies
involved, as well as a partitioned comparison of different results. Thus, a Schechter function was fitted to the completeness-corrected data given in Table 
\ref{otelo_lf_values} using a least-squares minimisation algorithm based on the Levenberg--Marquardt method.
The parameters obtained from the 
completeness-corrected LFs reported during recent years are summarised in Table \ref{otelo_lfs}.

\begin{figure*}[!htb]
\centering
\includegraphics[angle=0,width=0.8\textwidth]{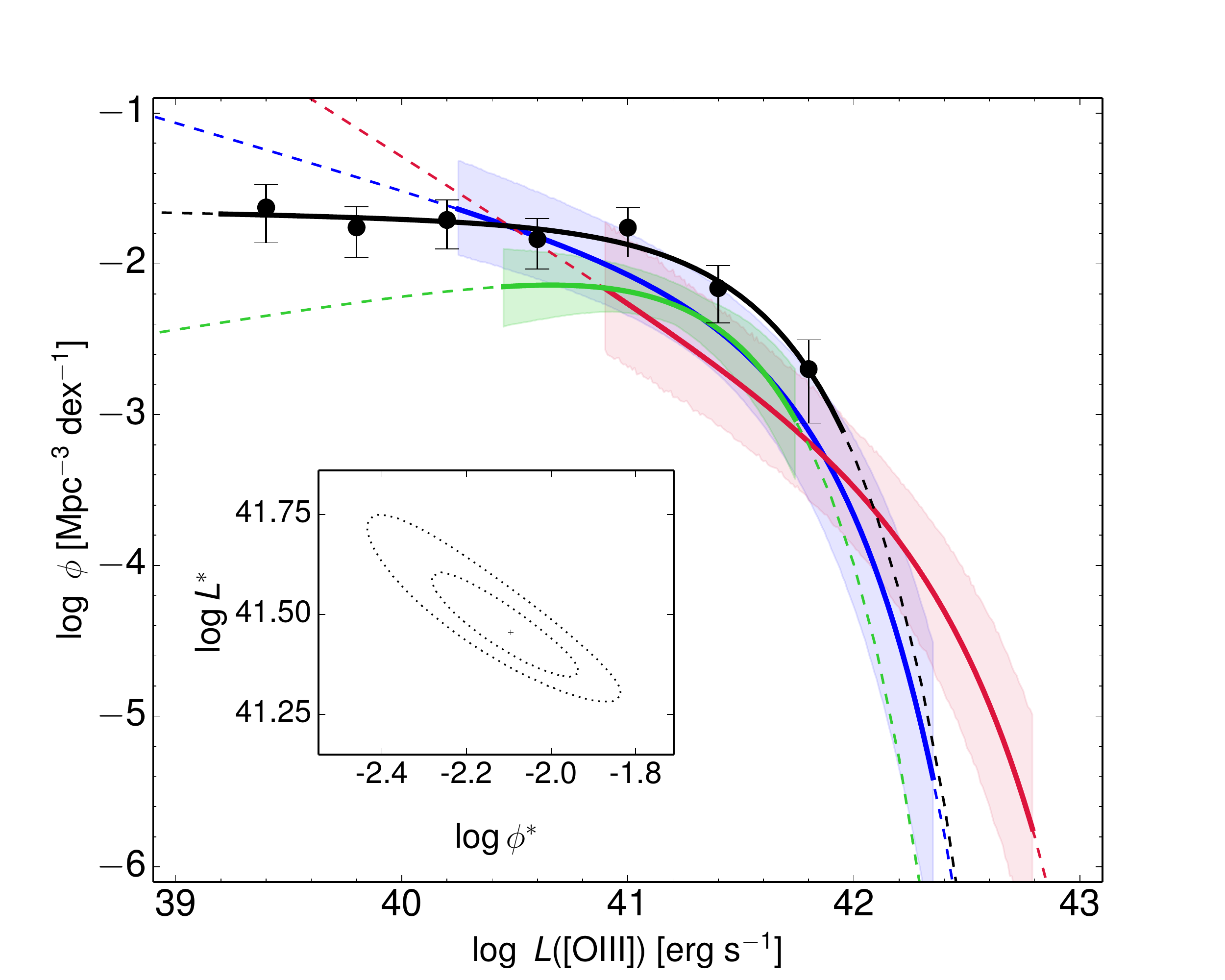}
\caption[]{Completeness-corrected \oiiis\ luminosity functions at $\langle$z$\rangle$$\sim$0.8 
from the recent literature and \otelo\ survey data. The observed (i.e.\ with no correction by dust attenuation) 
LFs of \cite{ly07}, \cite{drake13}, and \cite{hayashi18} are plotted in 
blue, green, and red, respectively. In each case, the 
solid line extends over the sampled luminosities in each survey, while the dashed line is the extrapolation of 
the corresponding best fit. 
The colour-shaded areas depict the propagated uncertainties of the cited LFs.
The black curve represents the best fit of the Schechter function to the \otelo\ LF(\oiiis) data (dots).
Error bars are the result of the LF uncertainty scheme per bin as detailed in Sect. \ref{sec:oiiiLFcalc}. 
The relevant data for this LF 
are given in Table \ref{otelo_lf_values}. The inset shows the 68\% and 90\% confidence contours for 
the \otelo\ data fit in the space of the parameters that show the largest uncertainties (see text).
}
\label{funlum}
\end{figure*}

Data from \cite{ly07} correspond to an NB survey of almost a dozen redshift windows between z=0.07 and 1.47 
explored in the Subaru Deep Field (SDF; \citealt{kashikawa04}). Using a maximum-likelihood analysis, 
\cite{drake13} derived an LF(\oiiis) from NB-selected ELSs in the Subaru/{\it XMM-Newton} Deep Field Survey 
(SXDS; \citealt{furusawa08}), employing the NB921 filter of the Suprime-Cam on the 8.2 m Subaru Telescope. The most 
recent estimate of the LF(\oiiis) at z$\sim$0.8 comes from the first Public Data Release of the
HSC-SSP \citep{hayashi18}. This is a comprehensive NB imaging survey of ELSs at
z $<$ 1.5 observed using the NB921 filter with the same telescope. Their LF parameters were obtained from
a combination of data over five extragalactic fields covering 16.2 deg$^2$ in total.
The resulting LF for our \oiiis\ ELS sample is shown in the Figure \ref{funlum} beside the plots of those 
obtained by the cited authors. 

\begin{table*}[!htb]
\vspace*{2mm}
\caption[Luminosity functions]{Best-fit Schechter parameters for \oiiis\ observed LFs extracted from the literature 
and the best fit of the \otelo\ LF(\oiiis) data given in Table \ref{otelo_lf_values}.}
\vspace*{-5mm}
\label{otelo_lfs}
\begin{center}
\addtolength{\tabcolsep}{-3.3pt}
\begin{tabular}{l c c c c c c c c c}
\hline   \\
Dataset & Method & Number of & Mean & Volume & $\log \phi^*$ & $\log L^*$ & $\alpha$ & $\log L_{\rm min}$ & $\log L_{\rm max}$ \\
\\[-0.9em]
 & & sources & redshift & [10$^3$ Mpc$^{3}$] & [Mpc$^{-3}$ dex$^{-1}$] & [erg s$^{-1}$] & & [erg s$^{-1}$] & [erg s$^{-1}$] \\
\\[-0.5em]
\hline     \\[-0.5em]
\cite{ly07} & NB & 662 & 0.84 & 43.65 & -2.54$\pm$0.15 & 41.53$\pm$0.11 & -1.44$\pm$0.09 & 40.25 & 42.36 \\ 
\\[-0.5em]

\cite{drake13} & NB & 910 & 0.83 & 123.50 & -2.25$^{+0.07}_{-0.09}$ & 41.28$^{+0.07}_{-0.08}$ & -0.76$^{+0.22}_{-0.19}$ & 40.45 & 41.75 \\ 
\\[-0.5em]

\cite{hayashi18} & NB & 6428 & 0.84 & 460-6300\tablefootmark{\,(a)} & -3.71$\pm$0.21 & 42.17$\pm$0.11 & -1.95$\pm$0.11 & 40.9 & 42.8 \\ 
\\[-0.5em]

This work & NB scan & 184 & 0.83 & 6.63 & -2.10$\pm$0.11 & 41.46$\pm$0.09 & -1.03$\pm$0.08 & 39.2 & 42.0 \\ 
\\[-0.5em]

\hline  
\end{tabular}
\end{center}
\tablefoot{
\tablefoottext{a}{value calculated using data from five extragalactic fields.} 
}
\end{table*}


\section{Discussion}
\label{discussion}

As stated above, uncertainties on the LF(\oiiis) due to CV (intrinsically induced by density fluctuations 
and galaxy clustering) are greater with increasing stellar mass and decreasing survey area for the same redshift range. 
Apart from the covariance of the characteristic luminosity $L^*$ with the LF normalisation $\phi^*$, errors 
on this latter parameter are also affected by the CV. All these uncertainties contribute to the inaccuracies of the LF bright-end 
predictions \citep{wolf03}. At the faint end, the completeness corrections become more important as the luminosity 
data reach the detection limits, whereas the CV effects are more limited. 

The largest relative uncertainty on the Schechter parameters obtained after the non-linear fitting of the \otelo\ LF(\oiiis) 
corresponds to the normalisation of the function ($\delta_r\, \phi^*$=25.5\%), followed by that of the characteristic 
luminosity ($\delta_r\, L^*$=18.2\%). The inset in Figure \ref{funlum} shows the strong correlation between both 
parameters. The relative error of the faint-end slope $\delta_r\, \alpha$ obtained is about 6.6\%.

Despite the difficulty in comparing Schechter-LFs because the parameters are more or less correlated, a general 
agreement can be observed between the LF estimates from previous studies and those of \otelo\ at 
$\log L$(\oiiis)\,$\approx$\,42. However, the \otelo\ LF(\oiiis) prediction in the vicinity 
of $L^*$ is $\sim$0.3 dex greater than those of \cite{ly07} and \cite{drake13}, and is about 0.6 dex greater than
the model of \cite{hayashi18}. Assuming a similar covariance of the LF normalisation with $L^*$ for all cases considered
here, we attribute such offsets to the differences between the normalisation parameters and their uncertainties, which as 
indicated above could be mainly related to the CV effects. Indeed, the fitted $\phi^*$ values tend to become smaller
as the order of magnitude of cosmic volume explored is greater. The volume sampled by \otelo\ is approximately 
7 and 1000 times smaller than those explored in previous studies. From the work of \cite{hayashi18} in particular, it is 
evident that the combination of five different samples of ELSs in a cosmic volume up to $\sim$10$^6$ Mpc$^{-3}$ not only allows mitigation 
of the small-number statistics affecting the LF bright-end, but also substantial reduction of the uncertainties 
linked to the CV, despite their significant relative errors on the parameters $\phi^*$ and $L^*$.
In summary, the volumes sampled by the authors cited here make their LF estimations more proficient than \otelo\ in 
sampling the bright side of the LF(\oiiis). Interestingly, the upper envelope of the LFs represented in Figure \ref{funlum} 
suggests that a double power-law function might better represents the bright side, as noted in recent investigations 
related to number statistics of ELGs \cite[see][and references therein]{comparat16}.

Even more remarkable is the behaviour of the LFs listed in Table \ref{otelo_lfs} at low luminosities. As 
a result of the combined effect of the line flux limit and EW$_{\rm obs}$ reached by \otelo, the LF reported 
in this work extends to luminosities that are about ten times fainter than the most sensitive observations of this kind made 
to date, yielding a direct constraint on the faint-end slope with a moderate relative error. This $\alpha$ value is 
shallower than those found by \cite{ly07} and \cite{hayashi18}, but steeper than that reported by \cite{drake13}; all of 
them resulting from NB data obtained with the same instrumental setup, but surveying different sky fields. As established
in Section \ref{sec:completeness}, a careful completeness correction of the observed \oiiis\ LF (Table \ref{otelo_lf_values}) 
based on well-informed simulations was performed before the Schechter function fitting. The mean uncertainty on
the CCF is not sufficient to explain these differences. Below, after an additional impact analysis of all LFs regarded in this work,
we explore some possibilities and physical scenarios that could help to account for different 
$\alpha$ values.

To further contrast the outcomes of the \oiiis\ LFs, we calculate the luminosity density 
$\mathfrak{L}({\rm [OIII]})$ and the space density $\mathfrak{N}({\rm [OIII]})$ of \oiiis\ ELSs, after
a numerical integration over the 39 $\leqslant$ $\log L({\rm [OIII]})$ [erg s$^{-1}$] $\leqslant$ 43 luminosity
range, which is roughly the same as that represented in the horizontal axis of Figure \ref{funlum}. An  estimation of the error on 
each integration was obtained from Monte Carlo analysis with 10\,000 realisations, assuming a normal 
distribution of the errors of the Schechter parameters given in Table \ref{lfs_stats}. In
Table \ref{lfs_stats} we show the resulting densities and their errors 
corresponding to each survey, as well as the mean values ($\mu$) and the standard deviations ($\sigma_{\mu}$) 
of these density calculations. 

As expected, and despite the discrepancies between LF parameters, the luminosity densities obtained are relatively 
self-consistent. Indeed, the standard deviation of the mean luminosity density is comparable to each uncertainty 
of $\mathfrak{L}({\rm [OIII]})$. In contrast, the number densities estimated over the same luminosity range 
show mutual divergences by a factor of between 3 and approximately 20, precisely because of the dispersion in the behaviour 
of the predicted or observed LFs on the faint side. Therefore, the faint-end slope of these LFs has a limited impact 
on the estimation of $\mathfrak{L}({\rm [OIII]})$ \citep[see also][]{drake13}, but a powerful effect on the 
number density of ELSs.

\newcommand*{\myalign}[2]{\multicolumn{1}{#1}{#2}}

\begin{table}[ht]
\vspace*{2mm}
\caption[Luminosity functions - statistics]{Luminosity $\mathfrak{L}$ and number $\mathfrak{N}$ densities of \oiiis\ 
ELSs calculated from the numerical integration of the observed LFs given in Table \ref{otelo_lfs}, over 
the 39 $\leqslant$ $\log L({\rm [OIII]})$ [erg s$^{-1}$] $\leqslant$ 43 luminosity range, as described in the main text. The mean 
values ($\mu$) and the standard deviations ($\sigma_{\mu}$) of these density parameters are given in the rows at 
the bottom.
}
\vspace*{-5mm}
\label{lfs_stats}
\begin{center}
\begin{tabular}{l c c}
\hline   \\
Dataset & $\log \mathfrak{L}({\rm [OIII]})$ & $\log \mathfrak{N}({\rm [OIII]})$ \\
\\[-0.9em]
& [erg s$^{-1}$] & [Mpc$^{-3}]$ \\
\\[-0.5em]
\hline     \\[-0.5em]
\cite{ly07} & {39.18$\pm$0.20} & {-1.124$\pm$0.228} \\ 
\\[-0.5em]

\cite{drake13} & {39.00$\pm$0.12} & {-1.812$\pm$0.205} \\ 
\\[-0.5em]

\cite{hayashi18} & {39.22$\pm$0.30} & {-0.676$\pm$0.381} \\ 
\\[-0.5em]

This work & {39.37$\pm$0.14} & {-1.351$\pm$0.153} \\ 
\\[-0.5em]
\hline  

\\[-0.5em]
\myalign{c}{$\mu$} & {39.19$\pm$0.10} & {-1.241$\pm$0.128} \\
\\[-0.5em]
\myalign{c}{$\sigma_{\mu}$} & {0.13} & {0.410} \\
\\[-0.5em]

\hline  

\end{tabular}
\end{center}
\end{table}

From a causal analysis, there are computational, statistical, and physical effects that could contribute to 
the scatter of the faint-end slopes of the LFs of SFGs. For instance, the analysis of the LF(UV) at z$\approx$2 
reported by \cite{parsa16} suggests that a robust faint-end slope estimate requires good sampling of the LF extending as far below $\log L^*$ as possible.
These latter authors demonstrated that there exists a dependence of the $\alpha$ value in the best-fitting Schechter
LF on the limiting absolute magnitude down to which this fit is performed, meaning that
brighter limits tend to give steeper slopes. Alternatively, \cite{drake13} not only claim that the detection 
fraction of ELSs plays an important role in the determination of the faint-end slope, but that this value is sensitive 
to the adopted limit of EW for the same faintest limit magnitude of a typical NB survey, which is on the order of the 
filter width. Based on this claim, and taking into account the EW lower limit of \otelo\ ELS data, namely about 6 \AA, 
the faint-end slope provided in this work is more robust than that obtained from the analysis of classic NB data.

In addition, \cite{drake13} also found that values of $\alpha$ for 
a given redshift could differ significantly, depending on the different emission lines used for the LF estimate. 
In this sense, \cite{sobral13} examined the behaviour of $\alpha$ by consistently using \ha\ ELSs observed in 
different redshift slices at 0.4 $<$ \zred\ $<$ 2.23, and contrary to a steepening with redshift found
in other works, they reported a relatively non-evolving value $\alpha =  1.60\pm0.08$. In particular, 
\cite{ly07} argues that even with robust completeness correction, there is an intrinsic
redshift evolution of the faint end of the star-forming galaxies towards a flatter slope. These latter authors also suggest
that the faint end of the LF at 0.8 $\la$ z $\la$ 1.5 estimated from NB data could be affected to some degree by
improper identification of \oii\ and \oiiis\ ELSs (in their case, by less than 10\%). Thus, misidentified lines
could artificially contribute to increasing $\alpha$ values for a given redshift. As established in Section
\ref{sec:oiiicontamination}, the final sample of \oiiis\ ELS described in this work is not declaredly affected by this
kind of contamination.
 
Concerning the physical scenarios, \cite{veilleux05} suggest that the faint end of the LF flattens as an effect of the
galactic winds produced by photo-evaporation in dwarf galaxies. The relatively shallow potential
of such galaxies is in agreement with this hypothesis. 
On the other hand, there are alternative energetic mechanisms (AGNs and SNe) that could explain the ionised gas outflows and
feedback, each entailing galaxy cooling, intergalactic medium (IGM) enrichment, and star-formation 
suppression \citep{yuma17}. In
the particular case of  galaxies that dominate the $L\ <<\ L^*$ regime, not only are supernova-driven winds able to explain
the metallicities of dwarf galaxies, but  the UV and X-ray backgrounds might also
heat the IGM \citep[see][and references therein]{guo11}. In both cases, a reduction in star formation 
may be expected. 
For these reasons, \cite{gargiulo15} claim that among the possible mechanisms 
(within the framework of the $\Lambda$-CDM paradigm) that regulate the faint-end slope of the LF in SFGs,
both the rate of SNe (which depends on
galaxy mass; in particular, the core-collapse type) and the efficiency 
of SNe feedback in the bulge and in the disc are of particular relevance. These effects can be introduced as free 
parameters for example in the semi-analytic models of galaxy formation. In this sense, \otelo\ provides data 
for the fine tuning of these parameters.

Finally, from an evolutionary point of view, these phenomena would explain the flattening of the LFs of SFGs since
cosmic epochs similar to or earlier than those explored in this paper, and force the galaxy to evolve towards low-mass 
early types  (dEs) that populate both cluster \citep{aguerri16} and field \citep{sybilska18} environments in 
the Local Universe. Under this general assumption, a fraction of the galaxies that contribute to the \otelo\ 
LF(\oiiis) could be predecessors of local post-starbursts and dEs.  


\section{Summary}

The OTELO survey is a new NB imaging survey that covers a 7.5\arcmin $\times$ 7.4\arcmin\ ($\sim$0.015 square degree) 
area in the EGS field. Using the RTF of the OSIRIS instrument on the GTC, 
a spectral window of 230 \AA\ in width, centred at $\sim$9170 \AA, and reasonably free from strong sky emission lines\ 
was scanned with 36 evenly spaced (every 6 \AA)  slices, each of 12 \AA\  in width.  From this spectral tomography we obtained 
low-resolution pseudo-spectra of all the sources detected on the coaddition of the slices. These sources are 
listed in a custom multi-wavelength catalogue which was built using public broad-band ancillary data
gathered within the framework of the AEGIS Collaboration. This catalogue contains a total of 9862 sources at 
50\% completeness magnitude of 26.38, and first-hand photometric redshift information with an uncertainty $\delta\,$\zp\ 
better than 0.2 (1+\zp) is available for 6660 of them. Details of the survey strategy, data 
reduction, and  main products are provided in \oteloone. 

Here, we demonstrate the scientific potential of the \otelo\ survey in a specific case for
the selection of an ELS sample to unprecedentedly low limits in line luminosity.

For this purpose we first established a general procedure for ELS selection using \otelo\ data. From 
extensive simulations
using synthetic pseudo-spectra, we estimated the limits of the survey in terms of emission line width,
EW, and flux as a function of  detection probability. These simulations allow us to derive a completeness
correction for the LF determination and other survey biases. 

The selection procedure was applied to the science case of \oiii\ ELSs. A total of 541 preliminary 
ELSs in the redshift window around \zred=0.8 were examined. Of these, a total of 184 sources 
constitute the final \oiiis\ ELS sample and another 92 from 
this preliminary set were classified as \hb, \heiihb, \ni, or \oiiia\ ELS candidates (which are being
used for further papers). After inverse
deconvolution of their emission lines, accurate redshifts, line fluxes, and observed EWs were determined
for the final \oiiis\ ELS sample, which is distributed in the range 0.78 $\leqslant$ \zotelo\ $\leqslant$ 0.87, 
with a mean
$\langle {\rm z}\rangle$ = 0.83. The vast majority (85\%) of the morphologically classified \oiiis\ ELSs are 
disc-like sources, and 87\% of this sample have a stellar mass ${\rm M}_\star < 10^{10}$ M$_{\odot}$. 
In a complementary manner, almost three quarters of the ELSs 
of this sample are dwarf galaxies. A limiting line flux of $\sim$5 $\times$ 10$^{-19}$ erg s$^{-1}$ cm$^{2}$ 
and observed EWs as low as $\sim$6 \AA\ were measured, which can be adopted as characteristic values of 
the survey as a whole. This validates the use of the TFs and the NB-scan technique for finding 
faint populations of SFGs, and in our opinion puts \otelo\ in a competitive position as 
the most sensitive survey to date in terms of minimum line flux and EW.

Sampling a comoving volume of $\sim7\, \times\, 10^3$ Mpc$^{3}$, and after analysing the contribution of
the most important uncertainties, we obtain an observed (i.e.\ with no correction for dust attenuation) LF 
from the final \oiiis\ ELS sample that is ten times fainter than the faintest extreme reached by other surveys; its best fits delivered
the following Schechter parameters: $\log \phi^*=-2.10\pm0.11$, $\log L^*=41.46\pm0.09$, and 
$\alpha=-1.03\pm0.08$. This faint-end slope value provides a new constraint on the behaviour of the
LF(\oiiis) and the number density of \oiiis\ ELSs at reddhift \zred=0.83, despite the agreement of the LF obtained 
with recent literature data in terms of the integrated \oiiis\ luminosity. In this sense, \otelo\ is complementary 
to other surveys with similar science goals and can help to put constraints on the parameters that regulate the 
statistics of dwarf SFGs in current models of galaxy evolution, specifically those dealing with feedback phenomena.

\begin{acknowledgements}

The Authors thank the anonymous referee for her/his feedback and useful suggestions,
and Terry Mahoney (at the IAC's Scientific Editorial Service) for his substantial
improvements of the manuscript.

AB thanks IAC researchers Bego\~na Garc\'ia Lorenzo, Julio Castro Almaz\'an, and 
Jos\'e A. Acosta Pulido for their useful comments.

This  work  was  supported  by  the  Spanish  Ministry  of  Economy  and
Competitiveness  (MINECO) under  the  grants
AYA2013\,-\,46724\,-\,P,
AYA2014\,-\,58861\,-\,C3\,-\,1\,-\,P,
AYA2014\,-\,58861\,-\,C3\,-\,2\,-\,P,
AYA2014\,-\,58861\,-\,C3\,-\,3\,-\,P,
AYA2016\,-\,75808\,-\,R,
AYA2016\,-\,75931\,-\,C2\,-\,1\,-\,P,
AYA2016\,-\,75931\,-\,C2\,-\,2\,-\,P and
MDM-2017-0737 (Unidad de Excelencia Mar\'ia de Maeztu, CAB).

This work was supported by the project Evolution of Galaxies, of reference 
AYA2017\,-\,88007\,-\,C3\,-\,1\,-\,P, within the "Programa estatal de fomento de la 
investigaci\'on cient\' ifica y t\' ecnica de excelencia del Plan Estatal de 
Investigaci\'on Cient\'ifica y T\'ecnica y de Innovaci\'on (2013-2016)" of the 
"Agencia Estatal de Investigaci\'on del Ministerio de Ciencia, Innovaci\'on y 
Universidades", and co-financed by the FEDER "Fondo Europeo de Desarrollo Regional".

JAdD thanks the Instituto de Astrof\'isica de Canarias (IAC) for its support
through the Programa de Excelencia Severo Ochoa and the Gobierno de Canarias for the Programa
de Talento Tricontinental grant.

MP acknowledges financial supports from the Ethiopian Space Science and Technology Institute
(ESSTI) under the Ethiopian Ministry of Innovation and Technology (MInT), and from the State
Agency for Research of the Spanish MCIU through the "Centre of Excellence Severo Ochoa" award
for the Instituto de Astrof\'isica de Andaluc\'ia (SEV-2017-0709). EJA acknowledges financial
support from the State Agency for Research of the Spanish MCIU through the "Centre of
Excellence Severo Ochoa" award for the Instituto de Astrof\'isica de Andaluc\'ia (SEV-2017-0709).

This article is based on observations made with the Gran Telescopio Canarias (GTC) at
 Roque de los Muchachos Observatory of the Instituto de Astrof\'isica de
Canarias on the island of La Palma.

This study makes use of data from AEGIS, a multi-wavelength sky survey conducted with the
Chandra, GALEX, Hubble, Keck, CFHT, MMT, Subaru, Palomar, Spitzer, VLA, and other telescopes, and is 
supported in part by the NSF, NASA, and the STFC.

Based  on  observations  obtained  with  MegaPrime/MegaCam,  a  joint  project  of the  CFHT  and
CEA/IRFU, at the Canada--France--Hawaii Telescope (CFHT) which is operated by the National
Research Council (NRC) of Canada, the Institut National des Science de l'Univers of the
Centre National de la Recherche Scientifique (CNRS) of France, and the University of
Hawaii.  This work is based in part on data products produced at Terapix available at
the Canadian Astronomy Data Centre as part of the Canada-France-Hawaii Telescope Legacy
Survey, a collaborative project of NRC and CNRS.

Based on observations obtained with WIRCam, a joint project of CFHT, Taiwan, Korea, Canada,
France, at the Canada--France--Hawaii Telescope (CFHT), which is operated by the National
Research Council (NRC) of Canada, the Institute National des Sciences de l'Univers of the
Centre National de la Recherche Scientifique of France, and the University of Hawaii.
This work is based in part on data products produced at TERAPIX, the WIRDS (WIRcam Deep
Survey) consortium, and the Canadian Astronomy Data Centre. This research was supported by
a grant from the Agence Nationale de la Recherche ANR-07-BLAN-0228.

\end{acknowledgements}

\bibliographystyle{aa} 
\bibliography{otelo2_v2} 

\begin{thebibliography}{}
\expandafter\ifx\csname natexlab\endcsname\relax\def\natexlab#1{#1}\fi

\bibitem[\'Alvarez et al.(1998)]{alvarez98} \'Alvarez, P., Rodr{\'{\i}}guez Espinosa, J.~M., \& S{\'a}nchez, F.\ 1998, New Astronomy Reviews, 42, 553

\bibitem[Aguerri(2016)]{aguerri16} Aguerri, J.~A.~L.\ 2016, \aap, 587, A111 

\bibitem[Arp(1966)]{arp66} Arp, H.\ 1966, \apjs, 14, 1 

\bibitem[Arnouts et al.(1999)]{arnouts99} Arnouts, S., Cristiani, S., Moscardini, L., et al.\ 1999, \mnras, 310, 540

\bibitem[Benitez et al.(2014)]{benitez14} Benitez, N., Dupke, R., Moles, M., et al.\ 2014, arXiv:1403.5237

\bibitem[Benson et al.(2003)]{benson03} Benson, A.~J., Bower, R.~G., Frenk, C.~S., et al.\ 2003, \apj, 599, 38

\bibitem[Bielby et al.(2012)]{bielby12} Bielby, R., Hudelot, P., McCracken, H.~J., et al.\ 2012, \aap, 545, A23 

\bibitem[Bongiovanni et al.(2019)]{OTELO1} Bongiovanni, {\'A}., Ram{\'o}n-P{\'e}rez, M., P{\'e}rez Garc{\'\i}a, A.~M., et al.\ 2019, \aap, 631, A9

\bibitem[Boselli et al.(2013)]{boselli14} Boselli, A., Hughes, T.~M., Cortese, L., Gavazzi, G., \& Buat, V.\ 2013, \aap, 550, A114 

\bibitem[Bothwell et al.(2011)]{bothwell11} Bothwell, M.~S., Kennicutt, R.~C., Johnson, B.~D., et al.\ 2011, \mnras, 415, 1815

\bibitem[Brough et al.(2011)]{brough11} Brough, S., Hopkins, A.~M., Sharp, R.~G., et al.\ 2011, \mnras, 413, 1236

\bibitem[Cedr{\'e}s et al.(2013)]{cedres13} Cedr{\'e}s, B., Beckman, J.~E., Bongiovanni, {\'A}., et al.\ 2013, \apjl, 765, L24 

\bibitem[Cepa et al.(2003)]{cepa03} Cepa, J., et al. \ 2003, SPIE, 4841, 1739

\bibitem[Chary \& Elbaz(2001)]{chary01} Chary, R., \& Elbaz, D.\ 2001, \apj, 556, 562

\bibitem[Comparat et al.(2016)]{comparat16} Comparat, J., Zhu, G., Gonzalez-Perez, V., et al.\ 2016, \mnras, 461, 1076 

\bibitem[Damjanov et al.(2018)]{damjanov18} Damjanov, I., Zahid, H.~J., Geller, M.~J., et al.\ 2018, \apjs, 234, 21

\bibitem[Danieli et al.(2018)]{danieli18} Danieli, S., van Dokkum, P., \& Conroy, C.\ 2018, \apj, 856, 69

\bibitem[Dark Energy Survey Collaboration et al.(2016)]{des16} Dark Energy Survey Collaboration, Abbott, T., Abdalla, F.~B., et al.\ 2016, \mnras, 460, 1270

\bibitem[Davies et al.(2018)]{davies18} Davies, L.~J.~M., Robotham, A.~S.~G., Driver, S.~P., et al.\ 2018, \mnras, 480, 768

\bibitem[Davis et al.(2007)]{davis07} Davis, M., Guhathakurta, P., Konidaris, N.~P., et al.\ 2007, \apjl, 660, L1 

\bibitem[Dawson et al.(2016)]{dawson16} Dawson, K.~S., Kneib, J.-P., Percival, W.~J., et al.\ 2016, \aj, 151, 44

\bibitem[De Lucia et al.(2014)]{delucia14} De Lucia, G., Muzzin, A., \& Weinmann, S.\ 2014, \nar, 62, 1 

\bibitem[Drake et al.(2013)]{drake13} Drake, A.~B., Simpson, C., Collins, C.~A., et al.\ 2013, \mnras, 433, 796 

\bibitem[Dressler et al.(2012)]{dressler12} Dressler, A., Spergel, D., Mountain, M., et al.\ 2012, arXiv:1210.7809

\bibitem[Driver et al.(2016)]{driver16} Driver, S.~P., Davies, L.~J., Meyer, M., et al.\ 2016, The Universe of Digital Sky Surveys, 42, 205

\bibitem[Driver et al.(2009)]{driver09} Driver, S.~P., Norberg, P., Baldry, I.~K., et al.\ 2009, Astronomy and Geophysics, 50, 5.12

\bibitem[Driver \& Robotham(2010)]{driverrobot10} Driver, S.~P., \& Robotham, A.~S.~G.\ 2010, \mnras, 407, 2131

\bibitem[Furusawa et al.(2008)]{furusawa08} Furusawa, H., Kosugi, G., Akiyama, M., et al.\ 2008, \apjs, 176, 1-18 

\bibitem[Gargiulo et al.(2015)]{gargiulo15} Gargiulo, I.~D., Cora, S.~A., Padilla, N.~D., et al.\ 2015, \mnras, 446, 3820 

\bibitem[Gil de Paz et al.(2003)]{gildepaz03} Gil de Paz, A., Madore, B.~F., \& Pevunova, O.\ 2003, \apjs, 147, 29

\bibitem[Grasshorn Gebhardt et al.(2019)]{grasshorn19} Grasshorn Gebhardt, H.~S., Jeong, D., Awan, H., et al.\ 2019, \apj, 876, 32

\bibitem[Guo et al.(2011)]{guo11} Guo, Q., White, S., Boylan-Kolchin, M., et al.\ 2011, \mnras, 413, 101 

\bibitem[Hayashi et al.(2018)]{hayashi18} Hayashi, M., Tanaka, M., Shimakawa, R., et al.\ 2018, \pasj, 70, S17 

\bibitem[Ilbert et al.(2006)]{ilbert06} Ilbert, O., Arnouts, S., McCracken, H.~J., et al.\ 2006, \aap, 457, 841

\bibitem[Kartaltepe et al.(2015)]{kartaltepe15} Kartaltepe, J.~S., Mozena, M., Kocevski, D., et al.\ 2015, \apjs, 221, 11 

\bibitem[Kashikawa et al.(2004)]{kashikawa04} Kashikawa, N., Shimasaku, K., Yasuda, N., et al.\ 2004, \pasj, 56, 1011

\bibitem[Klypin et al.(2015)]{klypin15} Klypin, A., Karachentsev, I., Makarov, D., et al.\ 2015, \mnras, 454, 1798

\bibitem[Laureijs et al.(2011)]{laureijs11} Laureijs, R., Amiaux, J., Arduini, S., et al.\ 2011, arXiv:1110.3193

\bibitem[Le F\`{e}vre et al.(2004)]{lefevre04} Le F\`{e}vre, O., Vettolani, G., Paltani, S., et al. 2004, A\& A, 428, 1043

\bibitem[Lilly et al.(2007)]{lilly07} Lilly, S.J., Le F\`{e}vre, O., Renzini, A., et al., 2007, ApJS, 172, 70

\bibitem[L{\'o}pez-Sanjuan et al.(2019)]{lopezsanjuan19} L{\'o}pez-Sanjuan, C., D{\'{\i}}az-Garc{\'{\i}}a, L.~A., Cenarro, A.~J., et al.\ 2019, \aap, 622, A51 

\bibitem[LSST Science Collaboration et al.(2009)]{lsst09} LSST Science Collaboration, Abell, P.~A., Allison, J., et al.\ 2009, arXiv:0912.0201

\bibitem[Lutz et al.(2011)]{lutz11} Lutz, D., Poglitsch, A., Altieri, B., et al.\ 2011, \aap, 532, A90

\bibitem[Ly et al.(2007)]{ly07} Ly, C., Malkan, M.~A., Kashikawa, N., et al.\ 2007, \apj, 657, 738 

\bibitem[McLure et al.(2018)]{mclure18} McLure, R.~J., Pentericci, L., Cimatti, A., et al.\ 2018, \mnras, 479, 25 

\bibitem[Moles et al.(2008)]{moles08} Moles, M., Benítez, N., Aguerri, J. A. L., et al., 2008, AJ, 136, 1325

\bibitem[Moster et al.(2011)]{moster11} Moster, B.~P., Somerville, R.~S., Newman, J.~A., \& Rix, H.-W.\ 2011, \apj, 731, 113 

\bibitem[Newman et al.(2013)]{newman13} Newman, J.~A., Cooper, M.~C., Davis, M., et al.\ 2013, \apjs, 208, 5 

\bibitem[Osterbrock \& Ferland(2006)]{osterbrock06} Osterbrock, D.~E., \& Ferland, G.~J.\ 2006, Astrophysics of gaseous nebulae and active galactic nuclei, 2nd.~ed.~by D.E.~Osterbrock and G.J.~Ferland.~Sausalito, CA: University Science Books, 2006

\bibitem[Parsa et al.(2016)]{parsa16} Parsa, S., Dunlop, J.~S., McLure, R.~J., \& Mortlock, A.\ 2016, \mnras, 456, 3194 

\bibitem[Pascual et al.(2007)]{pascual07} Pascual, S., Gallego, J., \& Zamorano, J.\ 2007, \pasp, 119, 30

\bibitem[P{\'e}rez-Gonz{\'a}lez et al.(2013)]{perezgonzalez13} P{\'e}rez-Gonz{\'a}lez, P.~G., Cava, A., Barro, G., et al.\ 2013, \apj, 762, 46 

\bibitem[S{\'a}nchez-Portal et al.(2015)]{sanchezportal15} S{\'a}nchez-Portal, M., Pintos-Castro, I., P{\'e}rez-Mart{\'{\i}}nez, R., et al.\ 2015, \aap, 578, A30

\bibitem[Sandage(1961)]{sandage61} Sandage, A.\ 1961, Washington: Carnegie Institution, 1961

\bibitem[Schechter(1976)]{schechter76} Schechter, P.\ 1976, \apj, 203, 297

\bibitem[Scodeggio et al.(2018)]{scodeggio18} Scodeggio, M., Guzzo, L., Garilli, B., et al.\ 2018, \aap, 609, A84 

\bibitem[Sobral et al.(2013)]{sobral13} Sobral, D., Smail, I., Best, P.~N., et al.\ 2013, \mnras, 428, 1128

\bibitem[Sobral et al.(2015)]{sobral15} Sobral, D., Matthee, J., Best, P.~N., et al.\ 2015, \mnras, 451, 2303

\bibitem[Somerville et al.(2012)]{somerville12} Somerville, R.~S., Gilmore, R.~C., Primack, J.~R., et al.\ 2012, \mnras, 423, 1992

\bibitem[Somerville, \& Primack(1999)]{somerville99} Somerville, R.~S., \& Primack, J.~R.\ 1999, \mnras, 310, 1087

\bibitem[Stasi{\'n}ska(2007)]{stasinska07} Stasi{\'n}ska, G.\ 2007, arXiv e-prints, arXiv:\ 0704.0348

\bibitem[Strauss et al.(2002)]{strauss02} Strauss, M.~A., Weinberg, D.~H., Lupton, R.~H., et al.\ 2002, \aj, 124, 1810 

\bibitem[Suzuki et al.(2016)]{suzuki16} Suzuki, T.~L., Kodama, T., Sobral, D., et al.\ 2016, \mnras, 462, 181

\bibitem[Sybilska et al.(2018)]{sybilska18} Sybilska, A., Kuntschner, H., van de Ven, G., et al.\ 2018, \mnras, 476, 4501 

\bibitem[Taylor et al.(2011)]{taylor11} Taylor, E.~N., Hopkins, A.~M., Baldry, I.~K., et al.\ 2011, \mnras, 418, 1587 

\bibitem[Veilleux et al.(2005)]{veilleux05} Veilleux, S., Cecil, G., \& Bland-Hawthorn, J.\ 2005, \araa, 43, 769

\bibitem[White, \& Frenk(1991)]{white91} White, S.~D.~M., \& Frenk, C.~S.\ 1991, \apj, 379, 52

\bibitem[Wolf et al.(2003)]{wolf03} Wolf, C., Meisenheimer, K., Rix, H.-W., et al.\ 2003, \aap, 401, 73

\bibitem[Yoachim et al.(2019)]{yoachim19} Yoachim, P., Graham, M., Bet, S., et al.\ 2019, \baas, 51, 303

\bibitem[Yuma et al.(2017)]{yuma17} Yuma, S., Ouchi, M., Drake, A.~B., et al.\ 2017, \apj, 841, 93 

\end{thebibliography}{}

\onecolumn

\appendix
\section{Examples of \oiiis\ ELS}
\begin{figure}[h]
\centering
  \begin{tabular}{cc}
    \includegraphics[angle=-90.,width=.45\textwidth]{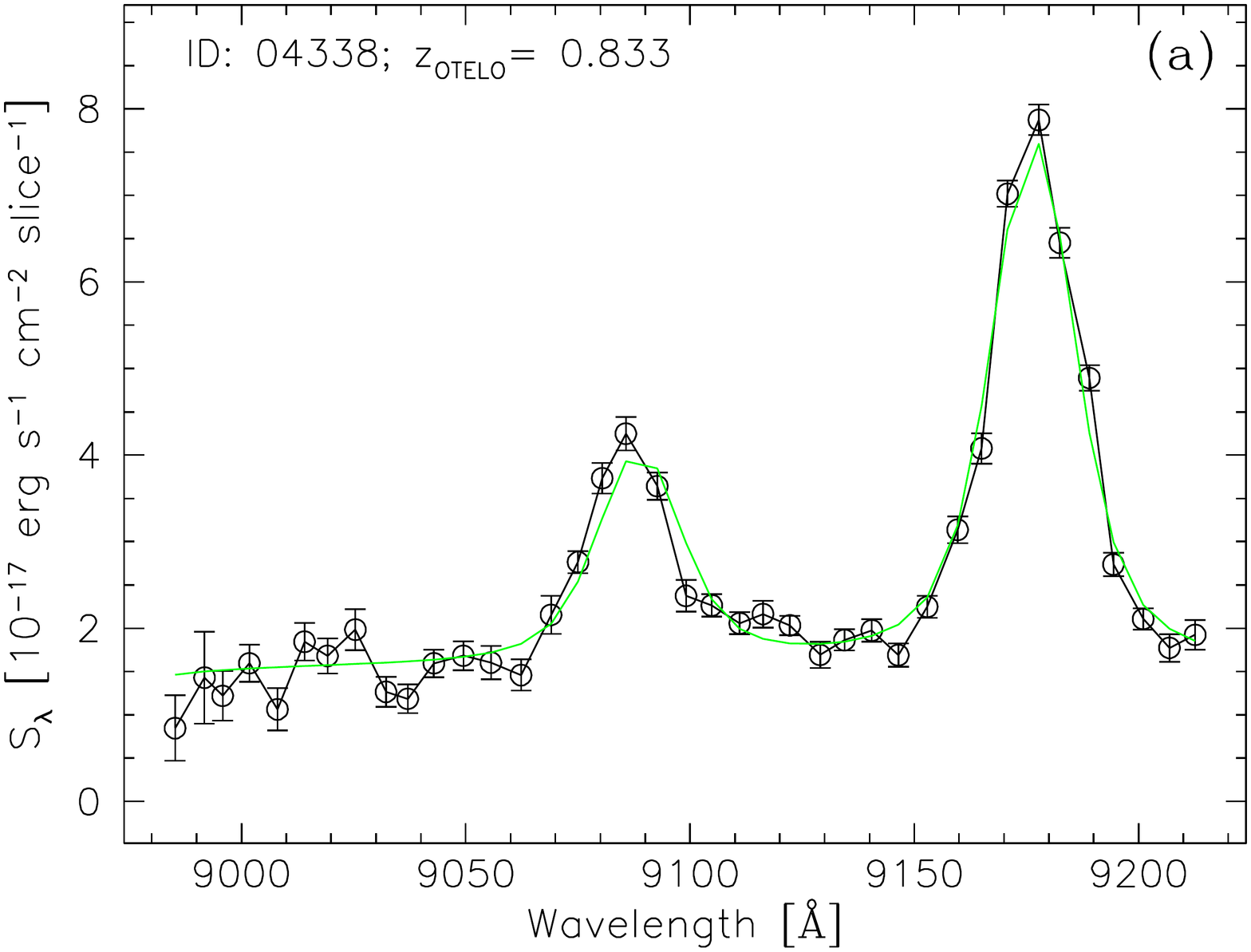} &
    \includegraphics[angle=-90.,width=.45\textwidth]{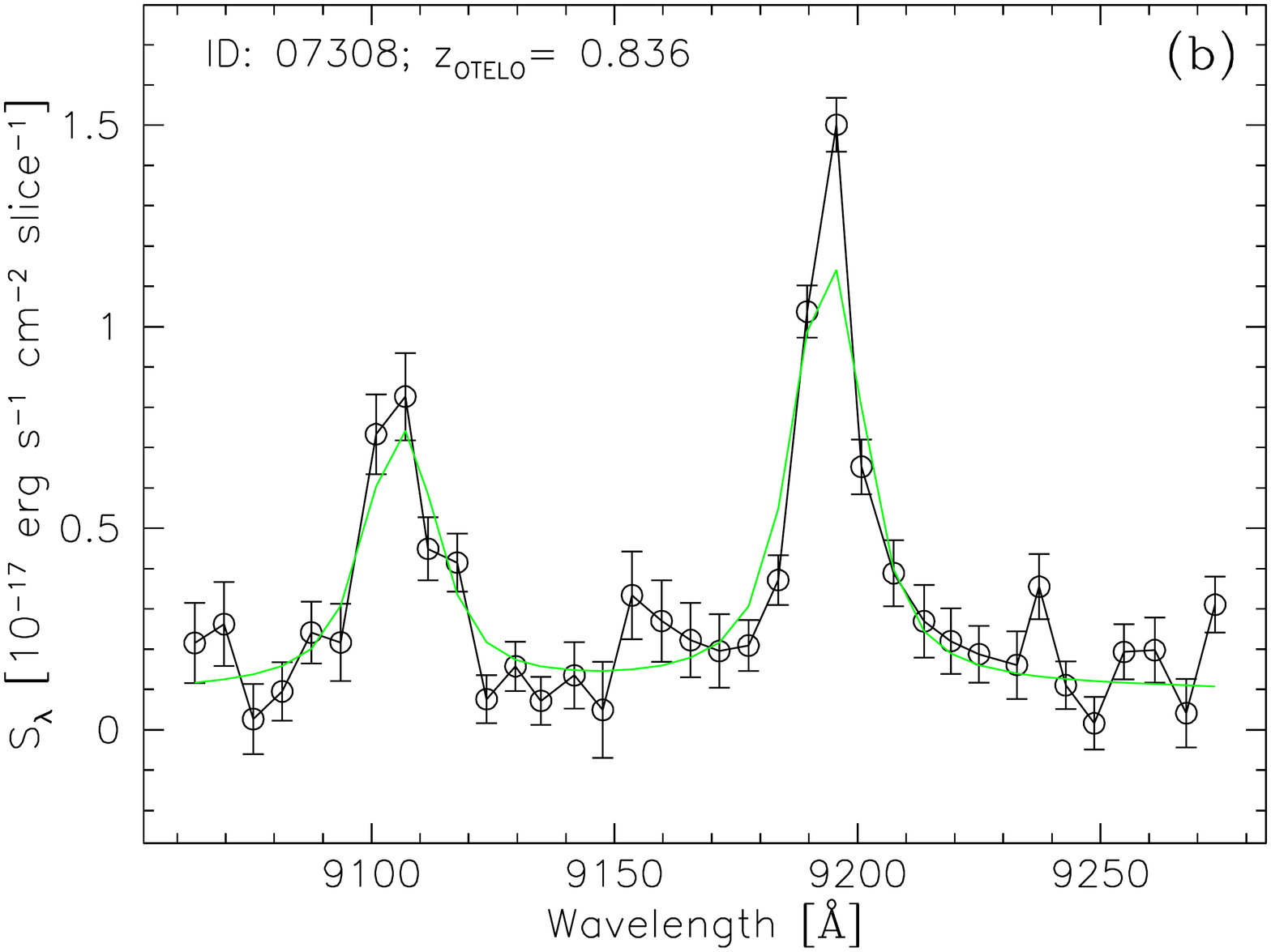} \\
    \includegraphics[angle=-90.,width=.45\textwidth]{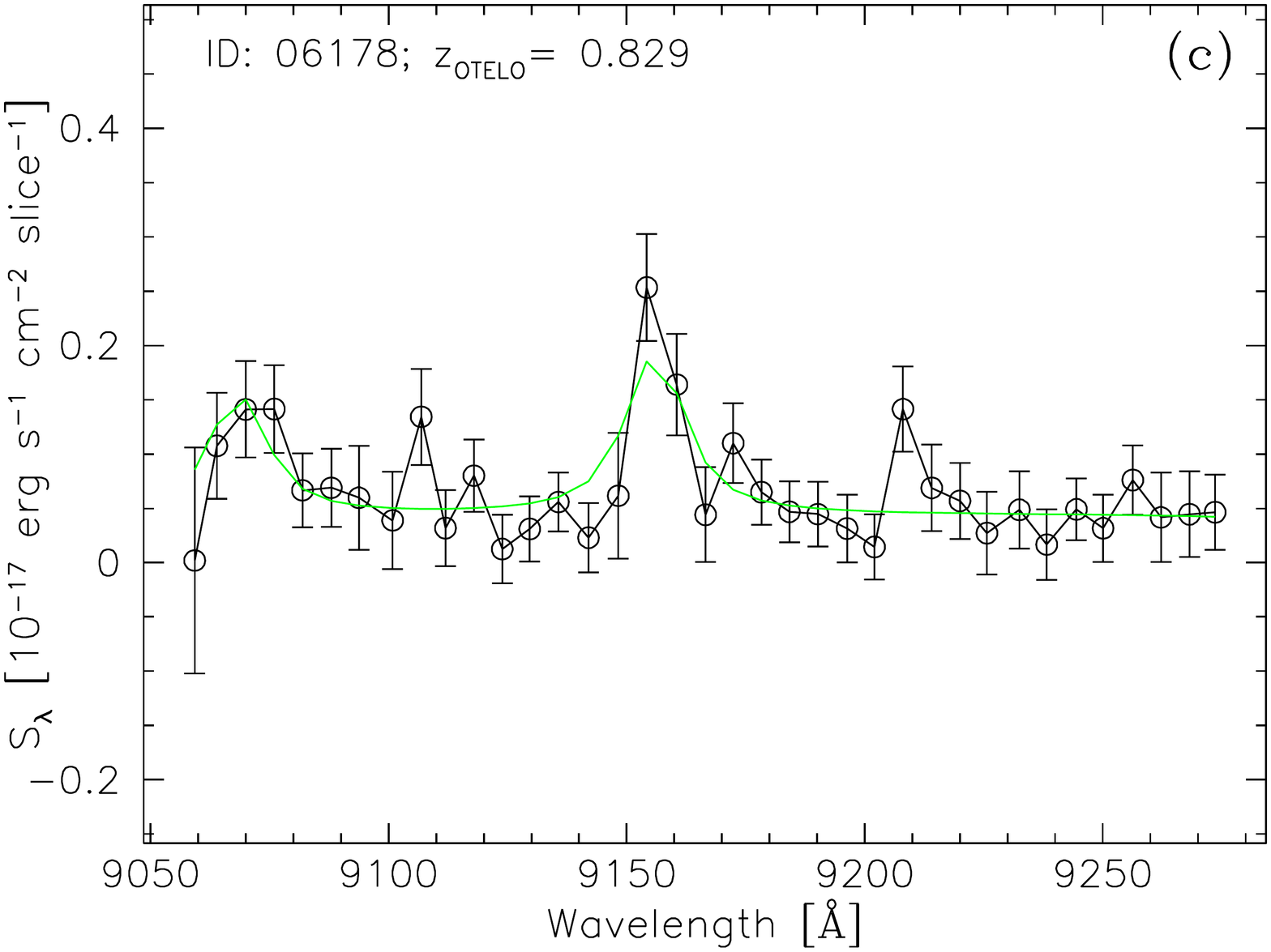} &
    \includegraphics[angle=-90.,width=.45\textwidth]{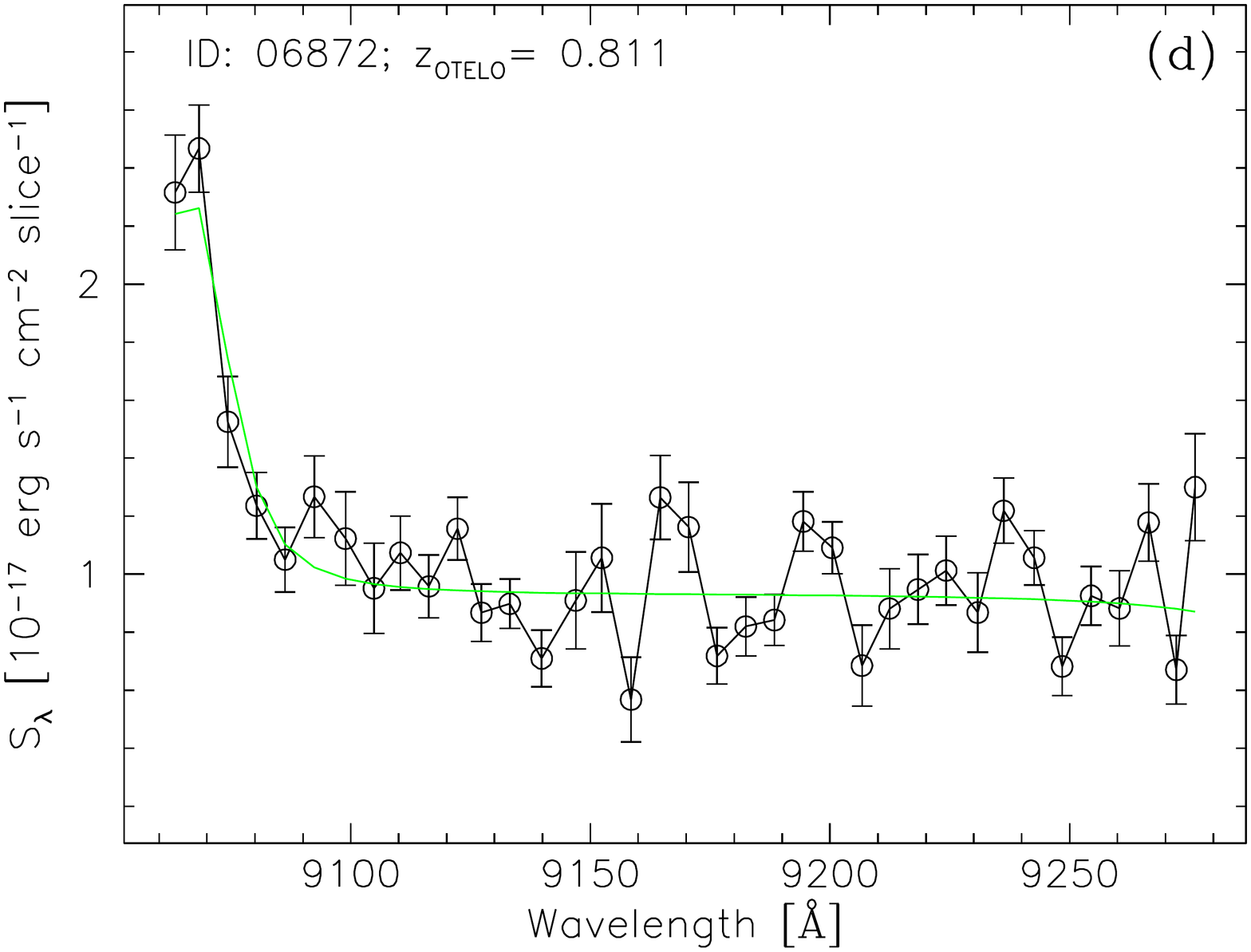} \\
    \includegraphics[angle=-90.,width=.45\textwidth]{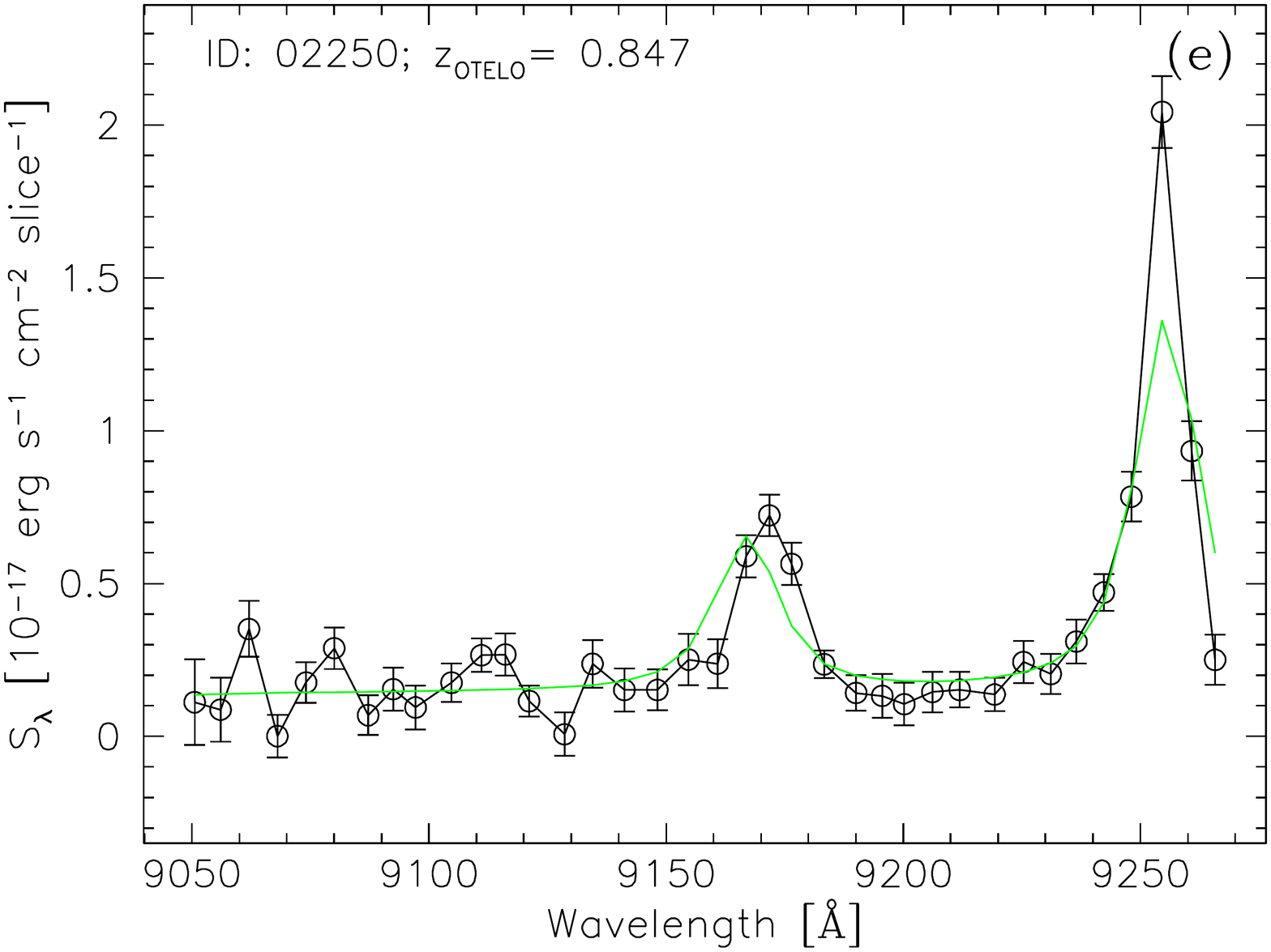} &
    \includegraphics[angle=-90.,width=.45\textwidth]{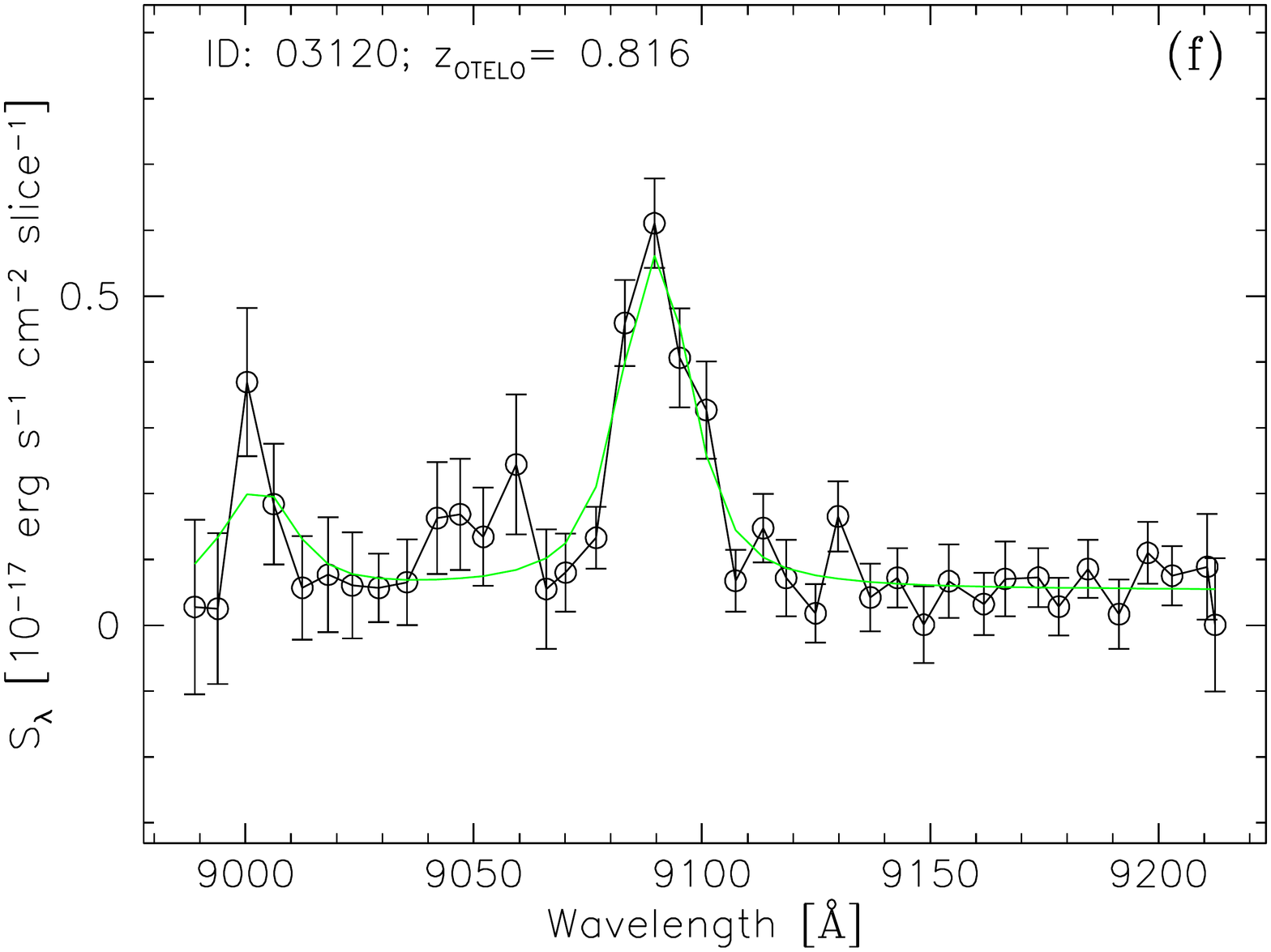} \\
  \end{tabular}
\captionsetup{margin=5pt,skip=20pt}
\caption{Examples of \otelo\ pseudo-spectra of \oiiis\ sources (open circles). The best theoretical pseudo-spectrum
obtained from the inverse deconvolution process described in Section \ref{sec:deconvolve} is represented by the green 
line. Panels (a), (b), and (c) show pseudo-spectra with a bright ({\tt ID}: 04338), a faint ({\tt ID}: 07308), and 
a very faint ({\tt ID}: 06178) continuum level, respectively. Panel (d)  
contains the pseudo-spectrum of a source ({\tt ID}: 06872) with the \oiiir\ line truncated on the
blue side of the survey spectral range. The pseudo-spectra in panels (e) and (f) -- {\tt ID}: 02250 and 03120 --
correspond to the \oiii\ sources unseen in ancillary data and only 
detected by visual examination of the pseudo-spectra (see Section \ref{sec:oiiieye} for details).
}
\label{PSexamples}
\end{figure}

\end{document}